\input harvmac
\input epsf
\noblackbox
%\draftmode

\newcount\figno
 \figno=0
 \def\fig#1#2#3{
\par\begingroup\parindent=0pt\leftskip=1cm\rightskip=1cm\parindent=0pt
 \baselineskip=5pt
 \global\advance\figno by 1
 \midinsert
 \epsfxsize=#3
 \centerline{\epsfbox{#2}}
 \vskip 10pt
 {\bf Fig.\ \the\figno: } #1\par
 \endinsert\endgroup\par
 }
 \def\figlabel#1{\xdef#1{\the\figno}}

\def\frac#1#2{{#1\over #2}}

\def\F{{ {\cal F} }}

\def\t{{ \tau }}
\def\vphi{{ { \varphi} }}

 %%% special math symbols
 \font\cmss=cmss10
 \font\cmsss=cmss10 at 7pt
 
 \def\inbar{\vrule height1.5ex width.4pt depth0pt}
 
 \def\IZ{\relax\ifmmode\mathchoice
 {\hbox{\cmss Z\kern-.4em Z}}{\hbox{\cmss Z\kern-.4em Z}}
 {\lower.9pt\hbox{\cmsss Z\kern-.4em Z}}
 {\lower1.2pt\hbox{\cmsss Z\kern-.4em Z}}\else{\cmss Z\kern-.4em Z}\fi}
 \def\IB{\relax{\rm I\kern-.18em B}}
 \def\IC{{\relax\hbox{$\inbar\kern-.3em{\rm C}$}}}
 \def\Ic{{\relax\hbox{$\inbar\kern-.22em{\rm c}$}}}
 \def\ID{\relax{\rm I\kern-.18em D}}
 \def\IE{\relax{\rm I\kern-.18em E}}
 \def\IF{\relax{\rm I\kern-.18em F}}
 \def\IG{\relax\hbox{$\inbar\kern-.3em{\rm G}$}}
 \def\IGa{\relax\hbox{${\rm I}\kern-.18em\Gamma$}}
 \def\IH{\relax{\rm I\kern-.18em H}}
 \def\II{\relax{\rm I\kern-.18em I}}
 \def\IK{\relax{\rm I\kern-.18em K}}
 \def\IP{\relax{\rm I\kern-.18em P}}
 \font\cmss=cmss10 \font\cmsss=cmss10 at 7pt
 \def\IR{\relax{\rm I\kern-.18em R}}

\def\sign{}

%%%%% more definitions %%%%%

\def\rN{N^{\rm orb}}
\def\rF{\F^{\rm orb}}
\def\iF{\F^{\infty}}

\def\dd{{\rm d}}
%%%%%%%%%%%%%%%%%%%%%%%%%%%%%%%%%%%%%%%%%%%%%%%%%%%%%%%%%%%%%%%%%%%%%%
\nref\ih{
  M.~Aganagic, R.~Dijkgraaf, A.~Klemm, M.~Mari\~no and C.~Vafa,
  ``Topological strings and integrable hierarchies,''
  Commun.\ Math.\ Phys.\  {\bf 261}, 451 (2006)
  [arXiv:hep-th/0312085].
  %%CITATION = HEP-TH 0312085;%%
}
\nref\AKMV{
M.~Aganagic, A.~Klemm, M.~Mari\~no and C.~Vafa,
  ``Matrix model as a mirror of Chern-Simons theory,''
  JHEP {\bf 0402}, 010 (2004)
  [arXiv:hep-th/0211098].
  %%CITATION = HEP-TH 0211098;%%
}
\nref\TV{
  M.~Aganagic, A.~Klemm, M.~Mari\~no and C.~Vafa,
  ``The topological vertex,''
  Commun.\ Math.\ Phys.\  {\bf 254}, 425 (2005)
  [arXiv:hep-th/0305132].
  %%CITATION = HEP-TH 0305132;%%
}
\nref\Allloop{
  M.~Aganagic, M.~Mari\~no and C.~Vafa,
  ``All loop topological string amplitudes from Chern-Simons theory,''
  Commun.\ Math.\ Phys.\  {\bf 247}, 467 (2004)
  [arXiv:hep-th/0206164].
  %%CITATION = HEP-TH 0206164;%%
}
\nref\BCOVI{
M.~Bershadsky, S.~Cecotti, H.~Ooguri and C.~Vafa, ``Holomorphic anomalies in
topological field theories,'' Nucl.\ Phys.\ B {\bf 405}, 279 (1993) [arXiv:hep-th/9302103].
  %%CITATION = HEP-TH 9302103;%%
}
\nref\BCOV{
  M.~Bershadsky, S.~Cecotti, H.~Ooguri and C.~Vafa,
  ``Kodaira-Spencer theory of gravity and exact results for quantum string amplitudes,''
  Commun.\ Math.\ Phys.\  {\bf 165}, 311 (1994)
  [arXiv:hep-th/9309140].
  %%CITATION = HEP-TH 9309140;%%
}
\nref\bryan{
  J.~Bryan et al, work in progress.
}
\nref\Candelas{
  P.~Candelas, X.~C.~De La Ossa, P.~S.~Green and L.~Parkes,
  ``A Pair Of Calabi-Yau Manifolds As An Exactly Soluble Superconformal
  Theory,''
  Nucl.\ Phys.\ B {\bf 359}, 21 (1991).
  %%CITATION = NUPHA,B359,21;%%
}
\nref\Cassels{
J.~W.~S.~Cassels, {\it Lectures on Elliptic Curves,} London Mathematical Society Student Texts {\bf 24}, Cambridge University Press, Cambridge, 1991.
}
\nref\ChiangTZ{
  T.~M.~Chiang, A.~Klemm, S.~T.~Yau and E.~Zaslow,
  ``Local mirror symmetry: Calculations and interpretations,''
  Adv.\ Theor.\ Math.\ Phys.\  {\bf 3}, 495 (1999)
  [arXiv:hep-th/9903053].
  %%CITATION = HEP-TH 9903053;%%
}
\nref\Connell{
I.~Connell, {\it Elliptic curve handbook,} McGill University, Montreal, 1996, available at http://www.math.mcgill.ca/connell/public/ECH1/.
}
\nref\deWit{
  B.~de Wit,
  ``N = 2 electric-magnetic duality in a chiral background,''
  Nucl.\ Phys.\ Proc.\ Suppl.\  {\bf 49}, 191 (1996)
  [arXiv:hep-th/9602060];
  %%CITATION = HEP-TH 9602060;%%
   B.~de Wit,
  ``N=2 symplectic reparametrizations in a chiral background,''
  Fortsch.\ Phys.\  {\bf 44}, 529 (1996)
  [arXiv:hep-th/9603191];
  %%CITATION = HEP-TH 9603191;%%
  B.~de Wit, G.~Lopes Cardoso, D.~Lust, T.~Mohaupt and S.~J.~Rey,
   ``Higher-order gravitational couplings and modular forms in N = 2,  D = 4
  heterotic string compactifications,''
  Nucl.\ Phys.\ B {\bf 481}, 353 (1996)
  [arXiv:hep-th/9607184].
  %%CITATION = HEP-TH 9607184;%%
}
\nref\MirrorSymmetryandEllipticCurves{
R.~Dijkgraaf, ``Mirror symmetry and elliptic curves," in {\it The moduli space of curves (Texel Island, 1994)}, Birkhäuser Boston, Boston, MA, 1995, pp. 149-163.
}
\nref\ng{
  R.~Dijkgraaf, S.~Gukov, A.~Neitzke and C.~Vafa,
  ``Topological M-theory as unification of form theories of gravity,''
  Adv.\ Theor.\ Math.\ Phys.\  {\bf 9}, 593 (2005)
  [arXiv:hep-th/0411073].
  %%CITATION = 00203,9,593;%%
}
\nref\DVV{
  R.~Dijkgraaf, E.~P.~Verlinde and M.~Vonk,
  ``On the partition sum of the NS five-brane,''
  arXiv:hep-th/0205281.
  %%CITATION = HEP-TH 0205281;%%
}

\nref\DouglasMoore{
  M.~R.~Douglas and G.~W.~Moore,
  ``D-branes, Quivers, and ALE Instantons,''
  arXiv:hep-th/9603167.
  %%CITATION = HEP-TH 9603167;%%
}
\nref\FarkasKra{
  H.~Farkas and I.~Kra, {\it Theta Constants, Riemann Surfaces and the Modular Group}, American Mathematical Society, 2001.
}
\nref\shatashvili{
  A.~A.~Gerasimov and S.~L.~Shatashvili,
  ``Towards integrability of topological strings. I: Three-forms on  Calabi-Yau
  manifolds,''
  JHEP {\bf 0411}, 074 (2004)
  [arXiv:hep-th/0409238].
  %%CITATION = JHEPA,0411,074;%%
}
\nref\Ghitza{
A.~Ghitza, ``An elementary introduction to Siegel modular forms," talk given at the Number Theory Seminar, University of Illinois at Urbana-Champaign,

 http://www.math.mcgill.ca/~ghitza/uiuc-siegel1.pdf.
}
\nref\HoriVafa{
  K.~Hori and C.~Vafa,
  ``Mirror symmetry,''
  arXiv:hep-th/0002222.
  %%CITATION = HEP-TH 0002222;%%
}
\nref\HK{
  M.~X.~Huang and A.~Klemm,
  ``Holomorphic anomaly in gauge theories and matrix models,''
  arXiv:hep-th/0605195.
  %%CITATION = HEP-TH 0605195;%%
}
\nref\KachruFV{
  S.~Kachru, A.~Klemm, W.~Lerche, P.~Mayr and C.~Vafa,
  ``Nonperturbative results on the point particle limit of N=2 heterotic string
  compactifications,''
  Nucl.\ Phys.\ B {\bf 459}, 537 (1996)
  [arXiv:hep-th/9508155].
  %%CITATION = HEP-TH 9508155;%%
}
\nref\kz{
M.~Kaneko and D.~B.~Zagier, ``A generalized Jacobi theta function and quasimodular forms," in {\it The moduli space of curves,} Progr. Math., 129, Birkhauser, Boston, MA, 1995, pp. 165-172.
}
\nref\KatzFH{
  S.~Katz, A.~Klemm and C.~Vafa,
  ``Geometric engineering of quantum field theories,''
  Nucl.\ Phys.\ B {\bf 497}, 173 (1997)
  [arXiv:hep-th/9609239].
  %%CITATION = HEP-TH 9609239;%%
}
\nref\klemmlerchetesisn{
  A.~Klemm, W.~Lerche and S.~Theisen,
  ``Nonperturbative effective actions of N=2 supersymmetric gauge theories,''
  Int.\ J.\ Mod.\ Phys.\ A {\bf 11}, 1929 (1996)
  [arXiv:hep-th/9505150].
  %%CITATION = HEP-TH 9505150;%%
}
\nref\klemmmarinotheisen{
  A.~Klemm, M.~Mari\~no and S.~Theisen,
  ``Gravitational corrections in supersymmetric gauge theory and matrix
  models,''
  JHEP {\bf 0303}, 051 (2003)
  [arXiv:hep-th/0211216].
  %%CITATION = HEP-TH 0211216;%%
}
\nref\KlemmZaslow{
A.~Klemm and E.~Zaslow,
  ``Local mirror symmetry at higher genus,''
  arXiv:hep-th/9906046.
  %%CITATION = HEP-TH 9906046;%%
}

\nref\Klingen{
H.~Klingen. {\it Introductory lectures on Siegel modular forms.} Vol. 20 of Cambridge
Studies in Advanced Mathematics, Cambridge University Press, Cambridge, 1990.
}
\nref\MarinoMoore{
  M.~Mari\~no and G.~W.~Moore,
  ``The Donaldson-Witten function for gauge groups of rank larger than one,''
  Commun.\ Math.\ Phys.\  {\bf 199}, 25 (1998)
  [arXiv:hep-th/9802185].
  %%CITATION = HEP-TH 9802185;%%
  M.~Mari\~no,
  ``The uses of Whitham hierarchies,''
  Prog.\ Theor.\ Phys.\ Suppl.\  {\bf 135}, 29 (1999)
  [arXiv:hep-th/9905053].
  %%CITATION = HEP-TH 9905053;%%
}
\nref\MW{
G.~W.~Moore and E.~Witten, ``Integration over the
u-plane in Donaldson theory,''  Adv.\ Theor.\ Math.\ Phys.\  {\bf 1}, 298 (1998)
[arXiv:hep-th/9709193].
  %%CITATION = HEP-TH 9709193;%%
}
\nref\nekrasov{
  N.~A.~Nekrasov,
  ``Seiberg-Witten prepotential from instanton counting,''
  Adv.\ Theor.\ Math.\ Phys.\  {\bf 7}, 831 (2004)
  [arXiv:hep-th/0206161].
  %%CITATION = HEP-TH 0206161;%%
}
\nref\NekrasovRJ{
  N.~Nekrasov and A.~Okounkov,
  ``Seiberg-Witten theory and random partitions,''
  arXiv:hep-th/0306238.
  %%CITATION = HEP-TH 0306238;%%
}
\nref\nekr{
  N.~Nekrasov,
  ``Z-theory: Chasing M-F-Theory,''
  Comptes Rendus Physique {\bf 6}, 261 (2005).
  %%CITATION = CRPOB,6,261;%%
}

\nref\yongbin{
 Y.-B.~Ruan et al., work in progress.
}
\nref\SW{
  N.~Seiberg and E.~Witten,
  ``Electric - magnetic duality, monopole condensation, and confinement in N=2
  supersymmetric Yang-Mills theory,''
  Nucl.\ Phys.\ B {\bf 426}, 19 (1994)
  [Erratum-ibid.\ B {\bf 430}, 485 (1994)]
  [arXiv:hep-th/9407087].
  %%CITATION = HEP-TH 9407087;%%
}
\nref\SWtwo{
  N.~Seiberg and E.~Witten,
  ``Monopoles, duality and chiral symmetry breaking in N=2 supersymmetric
  QCD,''
  Nucl.\ Phys.\ B {\bf 431}, 484 (1994)
  [arXiv:hep-th/9408099].
  %%CITATION = HEP-TH 9408099;%%
}

\nref\quantumsymmetries{
  C.~Vafa,
  ``Quantum Symmetries Of String Vacua,''
  Mod.\ Phys.\ Lett.\ A {\bf 4}, 1615 (1989).
  %%CITATION = MPLAE,A4,1615;%%
}
\nref\orbifwithdiscretetorsion{
  C.~Vafa and E.~Witten,
  ``On orbifolds with discrete torsion,''
  J.\ Geom.\ Phys.\  {\bf 15}, 189 (1995)
  [arXiv:hep-th/9409188].
  %%CITATION = HEP-TH 9409188;%%
}
\nref\verlinde{E.P.Verlinde,
  ``Attractors and the holomorphic anomaly,''
  arXiv:hep-th/0412139.
  %%CITATION = HEP-TH/0412139;%%
}
\nref\wittenBI{
  E.~Witten,
  ``Quantum background independence in string theory,''
  arXiv:hep-th/9306122.
  %%CITATION = HEP-TH 9306122;%%
}

%%%%%%%%%%%%%%%%%%%%%%%%%%%%%%%%%%%%%%%%%%%%%%%%%%%%%%%%%%%%%%%%%%%%%%%%%
% Title
%%%%%%%%%%%%%%%%%%%%%%%%%%%%%%%%%%%%%%%%%%%%%%%%%%%%%%%%%%%%%%%%%%%%%%%%%
\Title{
  \vbox{\baselineskip12pt \hbox{hep-th/0607100}
%\hbox{UCB-PTH/05/XX}
  \vskip-.5in}
}{\vbox{
  \centerline{Topological Strings}
\smallskip
\centerline{and}
\smallskip
\centerline{(Almost) Modular Forms}
}}

\centerline{Mina Aganagic,$^{1}$ Vincent Bouchard,$^2$ Albrecht Klemm,$^3$}
\bigskip\medskip
\centerline{$^1$ \it University of California, Berkeley, CA 94720, USA}
\vskip .03in
\centerline{$^2$ \it Mathematical Sciences Research Institute, Berkeley, CA 94720, USA}
\vskip .03in
\centerline{$^3$ \it University of Wisconsin, Madison, WI 53706, USA}
\medskip
\medskip
\medskip
\vskip .5in

%\baselineskip14pt
%%%%%%%%%%%%%%%%%%%%%%%%%%%%%%%%%%%%%%%%%
% Abstract
%%%%%%%%%%%%%%%%%%%%%%%%%%%%%%%%%%%%%%%%%
%\noindent
\vskip .1in \centerline{\bf Abstract}
{The B-model topological string theory on a Calabi-Yau threefold $X$
has a symmetry group $\Gamma$, generated by monodromies of the periods
of $X$.  This acts on the topological string wave function in a
natural way, governed by the quantum mechanics of the phase space
$H^3(X)$.  We show that, depending on the choice of polarization, the
genus $g$ topological string amplitude is either a holomorphic
quasi-modular form or an almost holomorphic modular form of weight $0$
under $\Gamma$. Moreover, at each genus, certain combinations of genus
$g$ amplitudes are both modular and holomorphic.
%This provides a powerful formulation of certain aspects of the work
%of Bershadsky et. al. in terms of
%symmetries of the theory.
We illustrate this for the
local Calabi-Yau manifolds giving rise to
Seiberg-Witten gauge theories in four dimensions and local $\IP_2$ and
$\IP_1\times \IP_1$. As a byproduct, we also obtain a
simple way of relating the topological string amplitudes near different
points in the moduli space, which we use to give predictions
for Gromov-Witten invariants of the orbifold ${\,\IC^3/\IZ_3}.$
\smallskip
\Date{July 2006}

\newsec{Introduction}

Topological string theory has led to many insights in both physics
and mathematics. Physically, it computes non-perturbative F-terms
of effective supersymmetric gauge and gravity theories in string
compactifications. Moreover, many dualities of superstring theory
are better understood in terms of topological strings.
Mathematically, the A-model explores the symplectic geometry and can be written in
terms of Gromov-Witten, Donaldson-Thomas or Gopakumar-Vafa invariants,
while the mirror B-model depends on the complex structure deformations
and usually provides a more effective tool for calculations.

The topological string is well understood for non-compact toric Calabi-Yau
manifolds. For example, the B-model on all non-compact toric Calabi-Yau
manifolds was solved to all genera in \refs{\ih} using the
$W_{\infty}$ symmetries of the theory. Geometrically, the $W_{\infty}$
symmetries are the $\omega$-preserving diffeomorphisms of the
Calabi-Yau manifold, where $\omega$ is the $(3,0)$ holomorphic volume form.  By contrast,
for compact Calabi-Yau manifolds the genus expansion of the topological string
is much harder to compute and so far only known up to genus four in certain cases, for instance for the quintic Calabi-Yau threefold.
It is natural to think that understanding quantum symmetries of the
theory may hold the key in the compact case as well.

In this paper, we will not deal with the full diffeomorphism group,
but we will ask how does the finite subgroup $\Gamma$ of large,
$\omega$-preserving diffeomorphisms, constrain the amplitudes. In
other words, we ask: what can we learn from the study of the group
of symmetries $\Gamma$ generated by monodromies of the periods of
the Calabi-Yau? For this, we need to know how $\Gamma$
acts in the quantum theory. 
The remarkable fact about the topological string is that its partition 
function $Z = \exp(\sum_{g} g_s^{2g-2} {\cal F}_g)$ 
is a wave function in a Hilbert space obtained by quantizing $H_{3}(X)$, 
where $g_s^2$ plays the role of ${\hbar}$.\foot{This fact was also recently explored in \refs{\shatashvili,\ng,\nekrasov,\verlinde}.} Classically, $\Gamma$ acts on $H_{3}
(X)$ as a discrete subgroup of the group $Sp(2n, \IZ)$ of symmetries
that preserve the symplectic form, where $n={1\over 2} b_3(X)$. This
has a natural lift to the quantum theory.

The answer turns out to be beautiful.
Namely, the ${\cal F}_g$'s turn out to be (almost) modular forms of
${\Gamma}$. By ``(almost) modular form" we mean one of two things: a
form which is holomorphic, but quasi-modular (i.e. it transforms with
shifts), or a form which is modular, but not quite holomorphic.  By
studying monodromy transformations of the topological string partition
function in ``real polarization'', where $Z$ depends holomorphically on the moduli space, we find that it is a quasi-modular form
of $\Gamma$ of weight $0$. 
The symmetry transformations
under $\Gamma$ imply that the genus $g$ partition function ${\cal
F}_g$ is fixed $recursively$ in terms of $lower$ $genus$ $data$, up to
the addition of a holomorphic modular form.  Thus, 
modular invariance constrains 
the wave function, but does not determine it uniquely. The holomorphic modular form that is picked out by the topological string can be deduced
(at least in principle) by its behavior at the boundaries on the
moduli space.  On the other hand, if we consider the topological
string partition function in ``holomorphic polarization'', 
this turns out to be a
modular form of weight $0$, which
is $not$ holomorphic on the moduli space. While it fails to be holomorphic, it turns
out to be ``almost holomorphic'' in a precise sense. Moreover, it is
again determined recursively, up to the holomorphic modular
form. Thus, the price to pay for insisting on holomorphicity is that
the ${\cal F}_g$'s fail to be precisely modular, and the price of
modularity is failure of holomorphicity!

The recursive relations we obtain contain exactly the same information
as what was extracted in \BCOV\ from the holomorphic anomaly
equation. In \BCOV, through a beautiful study of topological sigma
models coupled to gravity, the authors extracted a set of equations
that the genus $g$ partition function ${\cal F}_g$ satisfies, expressing an
anomaly in holomorphicity of ${\cal F}_g$.
The equations turn out to fix ${\cal F}_g$ in terms of lower genus data,
up to an holomorphic function with a finite set of undetermined coefficients.
Here, we have formulated the solutions to the holomorphic anomaly equation
by exploiting the underlying symmetry of the theory.
In the context of \BCOV , solving the equations was laborious,
the particularly difficult part being
the construction of certain ``propagators''. From our perspective,
the propagators are simply the ``generators'' of (almost) modular
forms, that is the analogues of the second Eisenstein series of
$SL(2,\IZ)$ and its non-holomorphic counterpart! That a
reinterpretation of \BCOV\ in the language of (almost)
modular forms should exist was anticipated by R. Dijkgraaf in
\refs{\MirrorSymmetryandEllipticCurves}.
For local Calabi-Yau manifolds, the relevant modular forms are Siegel modular forms. In the compact Calabi-Yau manifold case, our formalism seems to predict the existence
of a new theory of modular forms of (subgroups of) $Sp(2n,\IZ)$, defined on spaces with Lorentzian signature (instead of the usual Siegel upper half-space).

The paper is structured as follows. In section 2, we describe the
B-model topological string theory, from a wave function perspective,
for both compact and non-compact target spaces.
 In section 3, we take a
first look at how the topological string wave function behaves under
the symmetry group $\Gamma$ generated by the monodromies. Then, we
give a more precise analysis of the resulting constraints on the wave
function in section 4. We also explain the close relationship between
the topological string amplitudes and (almost) modular forms in this
section. In the remaining sections we give examples of our formalism:
in section 5 we study $SU(N)$ Seiberg-Witten theory, in section 6
local $\IP^2$ --- where we also use the wave function formalism to extract the Gromov-Witten invariants of the orbifold ${\,\IC^3/\IZ_3}$, and in section 7 local $\IP^1 \times \IP^1$. To conclude our work, in section 8 we present some open questions, speculations and ideas for future research. Finally, 
Appendix A and B are
devoted to a review of essential facts and conventions about modular
forms, quasi-modular forms and Siegel modular forms.

\newsec{B-model and the Quantum Geometry of $H^3(X,{\IC})$}

The B-model topological string on a Calabi-Yau manifold $X$ can be
obtained by a particular topological twisting of the ``physical''
string theory, two-dimensional $(2,2)$ supersymmetric sigma model on
$X$ coupled to gravity.  The genus zero partition function of the
B-model ${\cal F}_0$ is determined by the variations of complex
structures on $X$.  The higher genus amplitudes ${\cal F}_{g>0}$ can
be thought of as quantizing this.  When $X$ has a mirror $Y$, this is
dual to the A-model topological string, which is the Gromov-Witten
theory of $Y$, obtained by an A-type twist of the physical
theory on $Y$. As is often the case, many properties of the theory become
transparent when the moduli of $X$ and $Y$ are allowed to vary, and
the global structure of the fibration of the theory over its moduli
space is considered. This is quite hard to do in the A-model directly,
but the mirror B-model is ideally suited for these types of questions.

\vskip 0.07 in
\subsec{Real Polarization}
\vskip 0.07 in

Let us first recall the classical geometry of $H^{3}(X, {\,\IC}) =
H^{3}(X,\IZ) \otimes\; {\IC}$. In the following, we will assume that
$X$ is a compact Calabi-Yau manifold, and later explain the
modifications that ensue in the non-compact, local case.

Choose a complex structure on $X$ by picking a particular 3-form
$\omega$ in $H^{3}(X,{\,\IC})$. Any other 3-form differing from this
by a multiplication by a non-zero complex number determines the same
complex structure. 
The set of $(3,0)$-forms is a line bundle ${\cal L}$ over the
moduli space ${\cal M}$ of complex structures. Given a symplectic
basis of $H_{3}(X, \IZ)$,
$$
A^I \cap B_J = \delta^{I}_{J},
$$
where $I,J =1,\ldots n,$ and $n={1\over 2} b_3(X)$, we can
parameterize the choices of complex structures by the periods
$$
x^I = \int_{A^I} \omega, \qquad p_I = \int_{B_I} \omega.
$$

The periods are not independent, but satisfy the special geometry
relation:
\eqn\lag{
p_I(x) = {\del\over \del x^I} {\cal F}_0 (x).
}
As is well known, ${\cal F}_0$ turns
out to be given in terms of the classical, genus zero, free energy of the topological
strings on $X$.

In the above, we picked a symplectic basis of $H_3$. Different choices
of symplectic basis differ by $Sp(2 n, \IZ)$ transformations:
\eqn\trans{
\eqalign{
{\tilde p}_{I}&={A_{I}}^J {p}_J + B_{IJ} {x}^J\cr
{\tilde x}^{I} &= C^{IJ}{p}_J + {D^{I}}_{J} {x}^J
}
}
where
$$
M=\pmatrix{A & B \cr C & D \cr} \in Sp(2n, \IZ).
$$
For future reference, note that the period matrix $\tau$,
defined by
$$
\tau_{IJ} = {\del \over \del x^J} p_I
$$
transforms as
\eqn\tt{
{\tilde \tau} = (A \tau + B)(C \tau +D)^{-1}.
}
For a discrete subgroup $\Gamma \subset Sp(2n, \IZ),$ the changes of
basis can be undone by picking a different 3-form $\omega.$
Conversely, we should identify the choices of complex structure that
are related by changes of basis of $H_3(X,{\IZ}).$
The $x$'s can be viewed as projective coordinates on the
Teichmuller space ${\cal T}$ of $X$, on which
$\Gamma$ acts as the mapping class group.
Consequently, the
space of inequivalent complex structures is 
$$
{\cal M} ={\cal T}/\Gamma.
$$
Generically, the moduli space ${\cal M}$ has
singularities in complex codimension one, and $\Gamma$ is generated by
monodromies around the singular loci.
%
%
%\vskip 0.07 in
%{\it{ii.  Quantum Geometry of $H^3(X)$; Real Polarization}}
%\vskip 0.07 in
%

It is natural to think of $H^{3}(X,\IZ)$ as a classical phase space,
with symplectic form,
$$
dx^I \wedge d p_I,
$$
and \lag\ as giving a lagrangian inside it. In fact, the analogy is
precise. As shown in \wittenBI ,
in the quantum theory
$x^I$ and $p_J$ become canonically conjugate
operators
\eqn\comm{
[\, p_I,x^J] = g_s^2\, \delta_{I}^{J}
}
where $g_s^2$ plays the role of $\hbar$,
and the topological string partition function
\eqn\pert{
Z(x^I)= g_s^{{\chi\over 24}-1}\exp{[\,\sum_{g=0}^{\infty} g_s^{2g-2}{\cal F}_g(x^I) ]},
}
where ${\cal F}_g$ is the genus $g$ free energy of the topological
string, becomes a wave function.

More precisely, the B-model topological string theory determines a
$particular$ state $|Z\rangle$ in the Hilbert space obtained by
quantizing $H^3(X, {\IZ})$. The wave function,
$$
\langle x^I|Z\rangle = Z(x^I)
$$
describes the topological string partition function in one, ``real''
polarization\foot{For us, $\omega$ naturally lives in the
complexification $H^{3}(X,\,\IC) = {\IC} \otimes H^{3}(X, \IR)$, so
``real'' polarization is a bit of a misnomer.} of $H^3(X)$.
The semi-classical, genus zero approximation to the topological string
wave function is determined by the classical geometry of $X$,
and the lagrangian \lag :
$$
p_I Z(x) =
g_s^2 {\del \over \del x^I} Z(x) \sim
({\del
\over \del x^I} {\cal F}_0)
\; Z(x) .
$$
The lagrangian does not determine the full quantum wave function.
In general, there are normal ordering ambiguities, and to resolve them, 
the full topological B-model string theory is needed.\foot{
Note that due to \comm , $g_s$ is a section of ${\cal L}$,
so that ${\cal F}_g$ is a section of ${\cal L}^{2-2g}$.
The full partition function is a section of
${\cal L}^{{\chi\over 24}-1}$, where $\chi$ is the Euler characteristic of the Calabi-Yau,
due to the prefactor.}

The partition function $Z$ implicitly depends on the choice of
symplectic basis. Classically, changes of basis $(p,x) \rightarrow
(\tilde{p},\tilde{x})$ which preserve the symplectic form are
canonical transformations of the phase space. For the transformation
in \trans , the corresponding generating function $S(x, {\tilde x})$
that satisfies
\eqn\defss{
dS = p_I dx^I - {\tilde p}_I d {\tilde x}^I
}
is given by\foot{Note that \defss\ only defines $S$ up to an addition
of a $constant$ on the moduli space. This ambiguity can be absorbed in
${\cal F}_1$, since only derivatives of it are physical anyhow.}
\eqn\gen{
S(x, \tilde{x}) = -{1 \over 2} (C^{-1} D)_{JK} x^{J} x^{K} + (C^{-1})_{JK}x^{J} {\tilde x}^K -{1\over 2}(AC^{-1})_{JK} {\tilde x}^J {\tilde x}^K.
}
This has an unambiguous lift to the quantum theory,
with the wave function transforming as\foot{It is important
to note that this makes sense only on the $large$
phase space, where the integral is over the $n$-dimensional
space spanned by the $x^I$'s. In particular, the choice of section of
${\cal L}$ does not enter.}
\eqn\wvtransf{
{\tilde Z}({\tilde x})= \int \,d{x} \;e^{-S(x,{\tilde x})/g_s^2}
\, {Z}({x}).
}

We should specify
the contour used to define \wvtransf; however,
as long as we work with the perturbative $g_s^2$ expansion of $Z(x)$,
the choice of contour does not enter. To make sense of \wvtransf\ then,
consider the saddle point expansion of the
integral.

Given ${\tilde x}^{I}$, the saddle point of the integral
$x^{I} = x^{I}_{cl}$ solves the classical
special geometry relations that follow from \trans\ :
$$
{\partial  S \over \partial x^{I} }|_{x_{cl}} = p_I(x_{cl}).
$$
Expanding around the saddle point, and putting
$$
x^{I} = x_{cl}^I + y^I,
$$
we can compute the integral over $y$ by summing Feynman diagrams where
\eqn\ip{
\Delta_{IJ} = -( {\tau} +C^{-1}D)_{IJ}
}
is the inverse propagator, and derivatives of ${\cal F}_g$,
\eqn\ver{
\del_{I_1}\ldots \del_{I_n} {\cal F}_g(x_{cl}),
}
the vertices. As a short hand we summarize the saddle point expansion by
%
%$$
%{\tilde {\cal F}}_1 =
%{\cal F}_1 - {1\over 2} \, \log \,\det\, ({-\Delta})=: {\cal F}_1 + \Gamma_1(\Delta),
%$$
%and
$$
{\tilde {\cal F}}_g ={\cal F}_g + \Gamma_g(\Delta^{IJ},~\del_{I_1}\ldots \del_{I_n}{\cal F}_{r<g}(x_{cl}))
$$
where $\Gamma_g(\Delta^{IJ},~\del_{I_1}\ldots \del_{I_n}{\cal F}_{r<g}(x_{cl}))$
is a functional that is determined by the Feynman rules in terms of the lower genus vertices $\del_{I_1}\ldots \del_{I_n} {\cal F}_r(x_{cl})$ for $r < g$ and the propagator $\Delta^{IJ}$. The latter is related to the inverse propagator ${\Delta}_{IJ}$ in \ip\ by
$\Delta^{IJ}\Delta_{JK} =\delta^{I}_{K}$. For example, at genus $1$ the functional is simply
$$
\Gamma_1(\Delta^{IJ}) = {1 \over 2} \log \det (-\Delta),
$$
where by $\Delta$ we mean the propagator $\Delta^{IJ}$ in matrix form. At genus two one has
\eqn\FGII{
\eqalign{
\Gamma_2(\Delta^{IJ},~\del_{I_1}\ldots \del_{I_n}{\cal F}_{r<2})=&
\;{\Delta}^{IJ}\;({1\over 2} {\del_I \del_J} {\cal F}_1  + \;{1\over 2}\; {\del_I}{\cal F}_1 \;
{\del_J} {\cal F}_1)\cr
+& \;  {\Delta}^{IJ} \Delta^{KL}( \;{1\over 2} \del_I {\cal F}_1
 \;  {\del_J \del_K \del_L} {\cal F}_0
+ \;{1\over 8} \; {\del_I \del_J \del_K \del_L} {\cal F}_0)\cr
+& \;{\Delta}^{IJ}
{\Delta}^{KL}
{\Delta}^{MN}({1\over 8} \;{\del_I \del_J \del_K}
{\cal F}_0 \;\,{\del_L \del_M \del_N} {\cal F}_0
\cr
&~~~~~~~~~~~~~~~~~~~~~+  \; {1\over 12} {\del_I \del_K \del_M} {\cal F}_0
\;\,{\del_J \del_L \del_N} {\cal F}_0),} }
where we suppressed the argument $x_{cl}$ for clarity.

It is easy to see from the path integral that this describes all
possible degenerations of a Riemann surface of genus $g$
to ``stable'' curves of lower genera, with $\Delta^{IJ}$ being the corresponding contact term, as shown in the figure below.
Stable here means that the conformal Killing vectors
were removed by adding punctures, so that
every genus zero component has at
least three punctures, and every genus one curve, one puncture.\foot{Note that in particular this implies that at each genus, the equations are independent of the choice of section of ${\cal L}$ we made, the left and the right hand side transforming in the same way.}

%%%%%%%%%%%%%%%%%%%%%%%%%%%%%%%%%%%%%%%%%
\vbox{
%\bigskip
\centerline{\epsfxsize 5.0truein\epsfbox{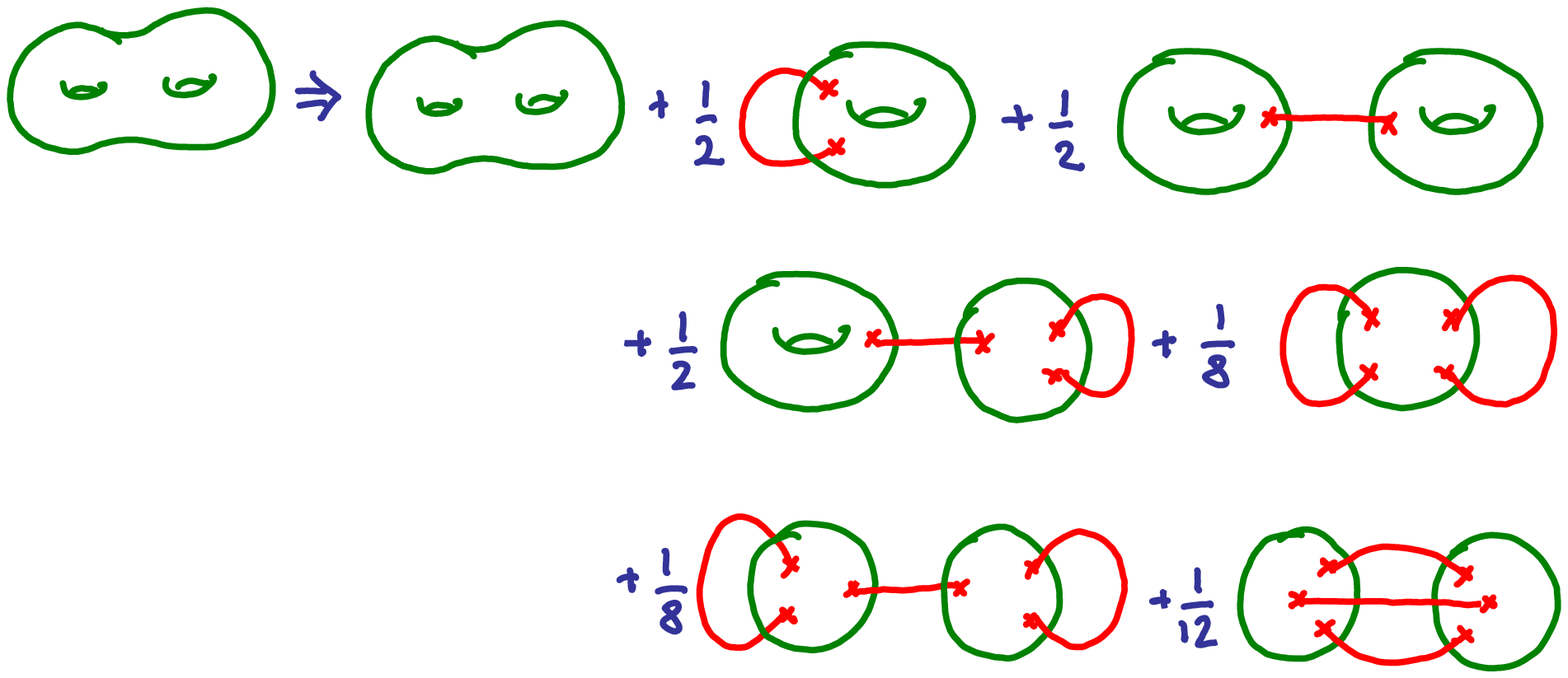}}
%\leftskip 2pc
%\rightskip 2pc
\noindent{\ninepoint
{\bf Fig. 1.} Pictorial representation of the Feynman expansion at genus $2$ in terms
of degenerations of Riemann surfaces.
{}}
\bigskip}
%%%%%%%%%%%%%%%%%%%%%%%%%%%%%

Mirror symmetry and Gromov-Witten theory picks out the real
polarization which is natural at large radius where instanton
corrections are suppressed, and where the classical geometry makes
sense. However, also by mirror symmetry, there is a larger family of
topological A-model theories which exist, though they may not have an
interpretation as counting curves.

For a generic element $M$ of $Sp(2n, \IZ)$, \wvtransf\ simply takes one
polarization into another. However, for $M$ in the mapping class group
$\Gamma \subset Sp(2n,\IZ)$, the transformation
\wvtransf\ should translate into a $constraint$ on ${\cal F}_g$, since
$\Gamma$ is a group of $symmetries$ of the theory.
We will explore the consequences of this in the rest of this paper.
\vskip 0.07 in
\subsec{Holomorphic Polarization}
\vskip 0.07 in
Instead of picking a symplectic basis of $H_3(X)$ to parameterize
the variations
of complex structure on $X$, we can choose a
fixed background complex structure
${\Omega} \in H^{3}(X,\, \IC)$, and use it to define
the Hodge decomposition of $H^{3}(X,{\IC})$:
$$
H^3 = H^{3,0}\oplus H^{2,1}\oplus H^{1,2}\oplus H^{0,3}.
$$
Here $\Omega$ is the unique $H^{3,0}$ form and the $D_i \Omega$'s
span the space of
$H^{2,1}$ forms, where
$D_i = \del_i - \del_i K$ and $K$ is the K\"ahler potential
$K = \log [i \int_X \Omega\wedge \bar{\Omega}]$.
This implies that:
\eqn\hol{
\omega = \vphi \,\Omega + z^i \, D_i \Omega+ {\bar z}^i
\,{\bar D}_i {\bar \Omega} + \bar{\vphi} \,{\bar \Omega},
}
where $(\vphi, z^i)$, and $(\bar{\varphi}, \bar{z}^i)$
become coordinates on the phase
space.\foot{Since $\omega$ for us does not live in
$H^{3}(X,\IR)$, but rather in $H^{3}(X, \, \IC)$, ${\bar \varphi}$ and
$\bar{z}^i$ are not honest complex conjugates of $\varphi$,
$z^i$.} Correspondingly we can express
$|Z\rangle$ as a wave function in holomorphic polarization
$$
\langle z^i, \vphi|Z\rangle = Z(z^i,\vphi).
$$

The topological string partition function $Z(z^i,\vphi)$ depends on the choice of background $\Omega$, and this dependence is not holomorphic. This is the holomorphic anomaly of
\refs{\BCOV}. One way to see this is through geometric quantization of
$H^{3}(X)$ in this polarization \wittenBI . We will take a different route, and exhibit this by exploring the canonical transformation from
real to holomorphic polarizations. Using special geometry relations
 it is easy to see that

$$\eqalign{
x^I& = \int_{A^I} \omega =  z^I+\; c.c\cr
p_I& = \int_{B_I} \omega =  \tau_{IJ}z^J  +\; c.c
}
$$
where we defined
$$
z^I=\vphi X^I + z^{i} D_i X^I
$$
in terms of
$$
X^I=\int_{A^I} \Omega, \qquad P_I = \int_{B_I} \Omega,
$$
and where
$$
\tau_{IJ} = {\partial \over \partial X^I} P_J.
$$
{}From this it easily follows that
$$
dp_I\wedge d x^I =
(\tau -{\bar \tau})_{IJ} d z^I \wedge d{\bar z}^J
$$
and hence the canonical transformation from $(x^I,p_I)$ to $(z^{I}, {\bar z}^I)$ is generated by
$$
d{\hat S}(x,z) = p_I dx^I + (\tau -{\bar \tau})_{IJ} {\bar z}^I d z^J.
$$
This corresponds to
$$
{\hat  S}(x,z) = {1\over 2} {\bar \tau}_{IJ} x^I x^J +
x^I(\tau - \bar{\tau})_{IJ} z^J - {1\over 2}
z^I(\tau - \bar{\tau})_{IJ} z^J + c,
$$
where $c$ is a constant, but which can now depend on the background.

In the quantum theory, this implies that the topological string partition function in the holomorphic polarization is related to that in real polarization by:
\eqn\delk{
{\hat Z}(z;t,{\bar t} )  = \int\, dx\; e^{- {\hat S}(x,z)/g_s^2} \, Z(x)
}
where $t^i$ are local coordinates on the moduli space, parameterizing
the choice of background, i.e. $X^I = X^I(t)$. Note that all the
background dependence of ${\hat Z}(z)$ comes from the kernel of ${\hat
S}$.\foot{In what follows, we will use hats to label quantities
which are not holomorphic.}
{}Let
\eqn\coeff{c(X,\bar X)=
- {\cal F}_1(X) -
{1\over 2} \log [\det(\tau - \bar \tau)](X,{\bar X})
- ({\chi\over 24} -1)\log(g_s),
}
where $\chi$ the Euler characteristic of the Calabi-Yau.

Consider now the perturbative expansion of the integral.
For simplicity, let us
pick
$$
\vphi=1, \qquad z^i=0,
$$
so that $z^{I}=X^I$.  The saddle point equation, which can be written
as\foot{We used here the special geometry relation $p_I =
\tau_{IJ}x^J.$}
$$
{(\bar \tau(X)- \tau(x_{cl}))}_{IJ}\,x_{cl}^J  + (\tau(X)-\bar{\tau}(\bar X))_{IJ}\,z^{J} =0,
$$
has then a simple solution,
$$
x^{I}_{cl} = X^I.
$$
Expanding around this solution,\foot{It should now be clear why \coeff\ is natural.  The
above normalization of the integral ensures that ${\hat Z}$ contains
no one loop term without insertions (the vanishing of genus zero terms
with zero, one and two insertions is automatic in the saddle point
expansion.)} 
we can compute the integral by summing Feynman diagrams where
\eqn\hp{
-(\tau(X) - \bar{\tau}({\bar X}))_{IJ}
}
%\eqn\hp{
%\hat{\Delta}_{IJ=}(\tau(X) - \bar{\tau}({\bar X}))_{IJ}
%}
is the inverse propagator, and derivatives of ${\cal F}_g$,
$$
\del_{I_1}\ldots \del_{I_n} {\cal F}_g(X),
$$
the vertices. That is, we get
\eqn\gtc{
{\hat{\cal F}}_g(t,\bar{t}) =  {\cal F}_g(X) + \Gamma_g\left(-\left((\tau-\bar{\tau})^{-1}\right)^{IJ},~
\del_{I_1}\ldots \del_{I_n} {\cal F}_{r<g} (X)\right)
}
where the properties of the functionals $\Gamma_g$ obtained by the
Feynman graph expansion have been discussed in the previous
section.

Finally, one can show \verlinde\ that $\hat Z$ satisfies the holomorphic anomaly equations of \BCOV . Differentiating the left and the right hand side of \delk\ the with respect to $\bar t$ we get
$$
{\del \over \del {\bar t}^{\bar i}} {\hat Z} =
[{g_s^2\over 2} {{\bar C}_{\bar i}}^{\;jk}
{\del^2 \over \del z^i \del z^j} + G_{{\bar i} j}
z^{j} {\del\over \del \varphi}]\; {\hat Z}
$$
In the above equation,
$C_{ijk}$ is the amplitude at genus zero with three punctures,
$G_{\bar i j}$ is the K\"ahler metric,
and ${{\bar C}_{\bar i}}^{\; jk} = e^{2 K}
{\bar C}_{{\bar i} {\bar j} {\bar k}} {G^{{\bar j} j} G^{\bar k k}}.$
It also satisfies the second holomorphic anomaly equation\foot{We used here the explicit form of $\hat F_1$ from \BCOV\ , from which follows that $\del_i \hat F_1 + ({\chi\over 24}-1) \del_i K = \del_i F_1 -{1\over 2}\del_i \log(\tau-\bar \tau)$.}
$$
 [{\del\over \del {t}^{\bar i}} +\del_i K (z^j
{\del\over \del z^j} - \varphi {\del\over \del \varphi})]{\hat Z} 
= [\varphi{\del\over \del z^i}
- \del_i{\hat {\cal F}}_1  -
({\chi\over 24}-1) \del_i K
-{1\over 2  g_s^2} {C}_{ijk}z^j z^k] {\hat Z}.
$$
The second anomaly equation implies that 
${\hat Z}$ has the form
$${\hat Z}(\varphi,z;t,{\bar t}) = \exp(\sum_{g,n} {1\over n!}\;g_s^{2g-2}
{\hat{\cal F}}^{(n)}_{g; i_1,\ldots i_n} z^{i_1}\ldots z^{i_n}
\varphi^{2-2g-n} - ({\chi\over 24} -1)\log \varphi)$$
where ${\hat {\cal F}}^{(n)}_{g; i_1,\ldots i_n}=D_{i_1}\ldots D_{i_n}
{\hat{\cal F}}_{g}$
for $2g-2+n>0$, and zero otherwise, for some ${\hat {\cal F}}_g$'s,
a fact that we will need later.

The holomorphic polarization, as explained in \refs{\BCOV, \wittenBI} is the
natural polarization of the topological string theory, in the
following sense. The topological string is obtained by twisting a
physical string on the Calabi-Yau at some point in the moduli space.
The physical string theory naturally depends not only on $X$, but also
on $\bar X$, so the space of physical theories is labeled by $(X,\bar
X).$ After twisting, it is natural to deform by purely topological
observables which are in one-to-one correspondence with the $h^{2,1}$ moduli
--- we have parameterized the resulting deformations by $z^i$
above. While one would naively expect the topological theory to
depend only on $z$, this fails and the theory depends on the background
$(X,{\bar X})$ that we used to define it as well.

\vskip 0.07 in
\subsec{ Local Calabi-Yau Manifolds}
\vskip 0.07 in

In the previous subsections we assumed that the Calabi-Yau $X$ is
compact. In this subsection we explain the modifications required in
the local case. We can derive the results of this section by viewing
the B-model on a local Calabi-Yau simply as a limit of the compact
one. This is the perspective that was taken in \refs{\ChiangTZ,\kz}.
%However, since then, significant
%progress has been made in understanding the B-model on local
%Calabi-Yau manifolds directly, without taking any limits \refs{
%\ih, \TV} --- in fact,
Since today, there is now far more known about the topological
string in the local than in the compact case, it is natural to work
directly in the language of local Calabi-Yau manifolds. For a string
theory on a non-compact Calabi-Yau manifold, gravity decouples. As a
consequence, the moduli space is governed by $rigid$ special
geometry, and not $local$ special geometry as in the compact
Calabi-Yau case. The partition functions are no longer sections of
powers of line bundle ${\cal L}$; the latter disappears altogether.

Consider the local Calabi-Yau manifold given by the equation
\eqn\LocalCY{
X\; : \;uw =  H(y,z)
}
in ${\;\IC}^4$.
This has a holomorphic three-form $\omega$ given by
\eqn\three{
\omega = {du\over u}\wedge dy \wedge dz.
}
The Calabi-Yau can be viewed as a $\,{\IC}^*$ fibration over the $y-z$
plane where a generic fiber is given by $uw = {\rm const}$.  It is easy to see that the 3-cycles on $X$ descend to 1-cycles on a Riemann surface $\Sigma$ given by
$$
\Sigma\; : \; 0= H(y,z) ,
$$
and, moreover, that the periods of the holomorphic three-form
$\omega$ on $X$ descend to the periods of a meromorphic 1-form $\lambda$
on $\Sigma$
$$
\int_{3-cycle} \omega = \int_{1-cycle} \lambda
$$
where
$$
\lambda = y dz.
$$

On a genus $g$ Riemann surface there
are $2g$
compact 1-cycles that form a symplectic basis,\foot{This is a slight over-simplification. Since the Riemann surface is non-compact, it can happen that one cannot find compact representatives of the homology satisfying this,
and that instead one has to work with $A^i \cap B_j =n^{i}_j,$
with $n^{i}_j$ integral. We will see examples of this in the later sections. Since it is very easy to see how this modifies the discussion of this section, we will not do this explicitly, but assume the simpler case for clarity of presentation.} $i=1,\ldots, g$,
$$
A^i \cap B_j =\delta^{i}_j.
$$
Let
$$
x^{i}
=
\int_{A^i} \lambda ,
\qquad p_{i} = \int_{B_i} \lambda ;
$$
the $x^{i}$'s are the $normalizable$ moduli of the Calabi-Yau manifold.
However, since the Calabi-Yau is non-compact, $H(y,z)$ may
depend on additional parameters which are {\it non-normalizable} complex
structure moduli $s^{\alpha}$.  Corresponding to these, there are
compact $3$-cycles $C_{\alpha}$ in $H_3(X)$ and $1$-cycles on
$\Sigma$ such that
$$
s^{\alpha} = \int_{C^{\alpha}} \lambda.
$$
But, since the homology dual cycles to the $C^{\alpha}$
are non-compact,
the metric on the moduli space along the corresponding directions
will not be normalizable. As a consequence, the
$s^{\alpha}$ are parameters of the model, not moduli.

This implies that the monodromy group
$\Gamma$ corresponds to elements of the form
\eqn\local{
\eqalign{
{\tilde p}_{i}&={A_{i}}^j {p}_j + B_{ij} {x}^j + E_{i\alpha} s^{\alpha}\cr
{\tilde x}^{i} &= C^{ij}{p}_j + {D^{i}}_{j} {x}^j + {F^{i}}_{\alpha} s^{\alpha}
}
}
where $s^{\alpha}$, being parameters which do not vary,
are monodromy invariant. Since $\Gamma$ preserves the symplectic form
$$
d x^{i} \wedge d p_i,
$$
we have that
$$
\pmatrix{A & B \cr C & D \cr} \in Sp(2g, \IZ).
$$
Note that, while $p_i$ and $x^j$ transform in a somewhat unconventional
way, the period matrix
$$
{\tau}_{ij} = {\del\over \del x^j}p_i
$$
transforms as usual:
$$
{\tilde \tau} = (A \tau+B)(C \tau + D)^{-1} .
$$
The corresponding generator of canonical transformations is easily found to be
\eqn\genloc{\eqalign{
S(x, \tilde{x}) =& -{1 \over 2} (C^{-1} D)_{jk} x^{j} x^{k} +
(C^{-1})_{jk}x^{j} {\tilde x}^k -
{1\over 2}(AC^{-1})_{jk} {\tilde x}^j {\tilde x}^k \cr
&+ C^{-1}_{ij} x^j F^{i}_{\alpha} s^{\alpha}
- E_{i \alpha} {\tilde x}^i s^{\alpha}.
}}
In the quantum theory, once again $x^i$ and $p_j$ are promoted to
operators with canonical commutation relations
$$
[x^i,p_j] = g_s^2\, \delta^{i}_{j}.
$$
The B-model determines a state $|Z\rangle$, and a wave function
$$
Z(x^i) = \langle x^i |Z \rangle.
$$
The wave function depends on the choice of real polarization,
the different polarization choices being related in the usual way:

\eqn\wvtloc{
{\tilde Z}({\tilde x})= \int \,d{x} \;e^{-S(x,{\tilde x})/g_s^2}
\, {Z}({x}).
}
Computing the path integral, in the saddle point expansion
around \local , we find that
\eqn\ipl{
\Delta_{ij} =-( {\tau} +C^{-1}D)_{ij}
}
is the inverse propagator, and derivatives of ${\cal F}_g$,
\eqn\ver{
\del_{i_1}\ldots \del_{i_n} {\cal F}_g(x_{cl}),
}
the vertices. This implies that
$$
{\tilde {\cal F}}_g =  {\cal F}_g + \Gamma_g({\Delta}^{ij},~{\del_{i_1}\ldots \del_{i_n}} {\cal F}_{r<g}(x_{cl}))\ ,
$$
where the propagator $\Delta^{ij}$ is related to \ipl\ by $\Delta^{ij}\Delta_{ij} =\delta^{i}_{k}$.

Now consider the holomorphic polarization. Once again, we pick a
background complex structure, this time by picking a meromorphic
1-form ${{\Lambda}}$ on $\Sigma$. Since we are not allowed to vary
the $C^{\alpha}$ periods, any other choice of complex structure
differing from this one by $normalizable$ deformations only
corresponds to picking a 1-form
$$
\lambda = {{\Lambda}} + z^{i} \, \del_i {{\Lambda}}+ {\bar z}^{i}
\,{\bar \del}_i {\bar{{\Lambda}}};
$$
here the ${\del_i \Lambda}$'s span a basis of holomorphic $(1,0)$-forms on
$\Sigma$ and correspond to infinitesimal deformations of complex
structures. This gives us a holomorphic set of coordinates on the phase space $(z^{i} , {\bar z}^i)$ which are canonically conjugate, and
allows us to write the wave function in the holomorphic polarization:
$$
{\hat Z}(z^{i})= \langle z^{i} | Z\rangle.
$$

We also need the relation between the two polarizations. Let
$$
X^i = \int_{A^i} \Lambda, \qquad P_i = \int_{B_i} \Lambda, \qquad s^{\alpha} = \int_{C^{\alpha}} \Lambda .
$$
It is easy to see that
$$
dx^i \wedge d p_i = (\tau_{ij} - \bar{\tau}_{ij})d {\bar Z}^i \wedge d{Z}^j
$$
where $\tau_{ij}(X) = \del P_i / \del X^j$ depends on the background and we put
$$
Z^{i} = z^{j} \del_j X^i .
$$
The corresponding canonical transformation is easily found:
$$
{\hat S}(x,z) ={1\over 2} \,
{\bar \tau}_{ij}\, (x-X)^{i} (x-X)^{j} + (\tau-\bar{\tau})_{ij}\, Z^{i} (x-X)^j-
{1\over 2}(\tau-\bar \tau)_{ij}\, Z^i Z^j + P_i x^i .
$$
The wave functions in holomorphic and real polarizations
are now simply related by
\eqn\relating{
{\hat Z}(z) = \int \, dx \, e^{-{\hat S}(x,z)/g_s^2}\, Z(x)
}
The saddle point equation reads
$$
\bar{\tau}({\bar X})_{ij}(x_{cl}-X)^j +
(\tau(X) - {\bar \tau}(\bar X))_{ij}Z^j
-(p-P)_i=0,
$$
and if we put $z^i=0$, which corresponds to $Z$ vanishing,
it has a simple solution:
$$
x_{cl}^i=X^i.
$$
Expanding around this, we get a Feynman graph expansion with
inverse propagator
$$
-(\tau(X)-\bar{\tau}({\bar X}))_{ij}
$$
and derivatives of ${\cal F}_g(X)$ as vertices. This gives the by
now familiar expansion relating the partition functions in
holomorphic and real polarizations:
\eqn\ftl{
{\hat {\cal F}}_g(t,{\bar t}) =  {\cal F}_g(t) +
\Gamma_g\left(-\left((\tau - \bar \tau)^{-1}\right)^{ij},~
\del_{i_1}\ldots\del_{i_n} {\cal F}_{r<g} (X)\right)\ .}

Before we go on, it is worth noting that the wave function in
holomorphic polarization satisfies a set of differential equations,
expressing the dependence of ${\hat Z}$ on the background --- the
$local$ $holomorphic$ $anomaly$ equations.  These can be derived easily
by differentiating both the left and the right hand side of \relating\
with respect to $\bar{t}$ (here, $t^i$ is the local coordinate
parameterizing the choice of background, $X=X(t)$).
%We will relegate the
%details to appendix A, and
This is straightforward, we state here only the answer:
\eqn\lhe{
{\del\over \del {\bar t}^{\bar i}} {\hat Z} = {1\over 2} \, g_s^2\,
\,{{\bar C}_{\bar i}}^{\;jk}
\,{\del^2 \over \del z^j \del z^k} {\hat Z}
}
where indices are raised by the inverse $g^{i \bar j}$ of the K\"ahler metric on the moduli space
$g_{i \bar j} = \del_i X^{k} (\tau - \bar \tau)_{k\ell}
 {\bar \del}_{\bar j} {\bar X}^{\ell}$.

In summary, apart from a few subtleties,
the quantum mechanics of the compact and local Calabi-Yau manifolds are
analogous. In the following section we will use the language of the
compact theory, but everything we will say will go over, without modifications,
to the non-compact case as well.

\newsec{A First Look at the $\Gamma$ Action}

In this section we take a first look at how topological string
amplitudes behave under monodromies.  On general grounds,
$\Gamma$ is a group of symmetries of the physical string theory.
This implies that the state $|Z\rangle$ in the Hilbert space that
the topological string partition function determines should be
$invariant$ under monodromies.
The associated wave functions, however, need not be.
By definition, the wave function in real
polarization requires a choice of symplectic basis of $H_3$ on which
$\Gamma $ acts nontrivially; thus, it cannot be monodromy invariant.
By contrast, the wave function in the holomorphic polarization
is the physical partition function. It is a well defined
function\foot{We are assuming a definite choice of gauge, throughout. Of course, changing the gauge, the amplitudes transform as sections of the apropriate powers of ${\cal L}$.} all over the moduli space; however, it is not holomorphic.

\subsec{The Wave Function in Real Polarization}

Given a symplectic basis $\{A^I, B_I\}$, $I=1,\ldots {n}$ of
$H_{3}(X,{\IZ})$, with $n = {1 \over 2} b_3$, a pick a
definite $3$-form $\omega$ in $H^{3}(X,{\IC})$.
The
topological string partition function determines a
wave
function
$$
Z(x^I) = \langle x^I | Z\rangle
$$
where
$$
x^I=  \int_{A^I} \omega,
$$
and a corresponding state $|Z\rangle$ in the Hilbert space obtained by
quantizing $H^{3}(X,{\IC})$. Having picked a definite section $\omega$
of the line bundle ${\cal L}$,
$x^I$'s and $Z(x^I)$ are at least locally, functions on the moduli space
$$
x^I = x^I(\psi).
$$
%%
%$$
%Z(x) = Z(x(\psi)),
%$$
where
the $n-1$ variables $\psi^i$
are some arbitrary local coordinates on ${\cal M}$.
For definiteness, we take here the Calabi-Yau manifold to be compact,
but everything carries over to the non-compact space as well,
the only real modification being that there the moduli space would have
dimension $n$, instead.

The moduli space ${\cal M}$ has singular loci in complex codimension $1$ around which
the cycles $A^I, B_J$ undergo monodromies in $\Gamma$.
As one goes around the singular locus,
by
sending $\psi$
$$
\psi \rightarrow \gamma\cdot\psi,
$$
for $\gamma$ an element of $\Gamma$,
the periods
%vector
%
%$$
%\Pi(\psi)\,\, = \,\pmatrix{\int_{B_I} \omega \cr \int_{A^I} \omega}
%$$
%
transform as
$$
\qquad\qquad\qquad\qquad\pmatrix{p_I \cr x^I}(\psi)\,\; \rightarrow \;
\pmatrix{p_I \cr x^I}(\gamma\cdot \psi)= \; M_{\gamma} \; \pmatrix{p_I\cr x^I}(\psi)\,
$$
where $M_{\gamma}$ is a symplectic matrix
corresponding to $\gamma$.

What happens in the quantum theory?
The monodromy group $\Gamma$ is a symmetry of the theory, so
the state $|Z\rangle$
determined by the topological string partition function
should be invariant under it:
$$
|Z\rangle \, \rightarrow \, |Z\rangle.
$$
The state $\langle x(\psi)|$, by contrast, is not invariant.
There are two ways to express what happens to $\langle x|$ under monodromies.
On the one hand, $x^I$ is a function of $\psi$, so we get a purely classical variation of the ket vector
$$
\langle  x(\psi)|\,\rightarrow\,
\langle x(\gamma\cdot \psi)|.
$$
But on the other hand, we have seen in section 2 that $any$
element  $M_\gamma\in Sp(2 n,{\IZ})$
acting classically on the period vector
has a unique lift to the quantum theory as an operator
$U_{\gamma}$ that acts on the Hilbert space. In particular,
$$
\langle  x ( \gamma \cdot \psi)| = \langle\, x (\psi)| \,U_\gamma .
$$
Putting these facts together implies that
$$
\langle x(\gamma\cdot \psi) |Z\rangle
= \langle x(\psi
) | \,U_\gamma\,|Z\rangle ,
$$
or, schematically in terms of wave functions,
\eqn\fund{
Z(x(\gamma\cdot \psi)) = \int \;  e^{S_{\gamma}} Z\,(x(\psi))
}
where
$\exp(S_\gamma)$ computes the corresponding matrix element of $U_\gamma$.
There is one such equation
for each monodromy transformation $g$ and its corresponding
element $M_{\gamma} \in \Gamma$.
Thus, the symmetry group $\Gamma$
imposes the constraints \fund\ on $Z$, one for each generator.

Using the results of section 2, equation \fund\ implies constraints on
the free energy, genus by genus.  For example, \fund\ implies that the
free energies satisfy\foot{It is important to emphasize
that this does not
depend on the choice of section either. We could
have written here simply
$x^I(\psi)=x^I$ and $x^I(\gamma \cdot \psi) =C^{IJ} p_J(x) + D^{I}_{\;J} x^J$.}
\eqn\modulartrans{
{{\cal F}}_g(x(\gamma\cdot \psi)) =  {\cal F}_g(x(\psi)) +
\Gamma_g\left(\Delta_{M_\gamma}^{IJ},~\del_{I_1}\ldots \del_{I_N}{\cal F}_{r<g}\right)}
with $\Delta_{M_{\gamma}}$ given by
\eqn\ipagain{
(\Delta_{M_{\gamma}})^{IJ} = -\left((\tau + C^{-1} D)^{-1}\right)^{IJ},
}
where
\eqn\monma{
M_{\gamma} = \pmatrix{A&B\cr C&D}.
}
%

%We can write the equations at arbitrary genus compactly as follows.
%Define the generating function of correlation functions
%
%$$
%{\cal W}(x(\psi)+y) = \,\sum_{g,n}\; {1\over n!}
%\; \del_{I_1}\ldots \del_{I_n} {\cal F}_g(x)  \;
%y^{I_1}\ldots y^{I_n}\;g_s^{2g-2}
%$$
%
%where the sum over $n$ runs from zero to infinity, except at genus
%zero and one, where it starts at $n=3$ and $n=1$, respectively.
%Then, we have
%
%$$
%e^{{\cal W}(x(M_\Gamma \,\psi))}
%= e^{-{g_s^2 \over 2 }(\Delta_{M_\Gamma})^{IJ}
%{\del\over \del y^I}{\del\over \del y^J}} \;
%e^{{\cal W}(x(\psi)+y)}
%$$
%
%where, after taking the derivatives, we set $y$ to zero.
%Below, we will show how to solve these equations and make the modular
%properties of ${\cal F}_g$ manifest.

To summarize, non-trivial monodromy (with $\det(C)\neq 0$)
around a point in the moduli space
corresponds to choosing $A$-cycles which are $not$ $well$ defined there, but instead transform by
$$
x^{I} \rightarrow C^{IJ} p_J + {D^{I}}_J x^{J}.
$$
This leads to an obstruction to analytic
continuation of the amplitudes all over the moduli space.
It also lead us to the notion of
``good variables'' in the moduli space, which are implicit in
Gromov-Witten computations: near a point in the moduli space, the ``good'' variables
are those with no non-trivial monodromy, meaning that $C^{IJ}=0$.

\subsec{Another Perspective}

Consider instead the wave function in holomorphic polarization.
Pick a
background complex structure $\Omega$, and write $\omega$ as in \hol\
$$
\omega = \vphi \,\Omega + z^i \, D_i \Omega+ {\bar z}^i
\,{\bar D}_i {\bar \Omega} + \bar{\vphi} \,{\bar \Omega}.
$$
Using $\vphi$ and $z^i$ as coordinates, we can write $|Z\rangle$ as a wave function in holomorphic polarization
$$
{\hat  Z}(\vphi, z^i)=\langle \vphi, z^i | Z\rangle
$$
Note that $z^i$ are coordinates on ${\cal M}$, centered at
$\Omega.$

How does $Z(\vphi, z^i)$ transform under $\Gamma$?  In real
polarization, the non-trivial transformation law of the wave function
came about from having to pick a basis of periods $\langle x^I|$,
which were not invariant under $\Gamma$. In writing down the wave
function in holomorphic polarization, that is in defining $\langle
\vphi, z^i|$, we made no reference to the periods, so ${\hat Z}(\vphi,
z^i)$ has to be invariant. There is another, independent reason why
this has to be so. Namely, $Z(\vphi, z^i)$ is the physical wave
function everywhere on ${\cal M}$ and as such, it better be well
defined everywhere!

We have seen above that the wave function in real polarization has
rather complicated monodromy transformations under $\Gamma$, while the
wave function in holomorphic polarization is invariant.
Since the two polarizations are related in a simple way, we could have derived the transformation properties of one from that of the other.
Consider for example the genus two amplitudes in \gtc\ for a compact
Calabi-Yau, and in \ftl\ for a non-compact one.
While on the left hand side $\hat{\F}_2$ is manifestly
invariant under $\Gamma$, on the right hand side all the ingredients have non-trivial monodromy transformations. In fact,
we have
\eqn\anom{
\left(( \t-\bar\t)^{-1}\right)^{IJ}\rightarrow
{(C\t +D)^{I}}_{K}\;{(C\t+D)^{J}}_L \left( ( \t-\bar\t)^{-1}\right)^{KL}-
C^{IL}{(C\t+D)^{J}}_L,
}
where $C$, $D$ enter $M_{\Gamma}$ as in \monma\ ,
%$$
%M_{\Gamma} = \pmatrix{A&B\cr C&D},
%$$
and analogously in the local case.
These quasi-modular transformations of $(\t-{\bar\t})^{-1}$ must
precisely cancel the transformations of the genus zero, one and
two amplitudes in real polarization. We will come back to this in the next section.

\newsec{Topological Strings and Modular Forms}

In the previous section we took a first look at how the topological
string partition functions transform under $\Gamma$.  In this section
we give a simple and precise description of how, and to which
extent, can the discrete symmetry group $\Gamma$ constrain the
topological string amplitudes. Along the way, we will discover a close
relationship of topological string partition functions and modular
forms.

On the one hand, we have seen in the previous sections that the partition function in
{\it holomorphic} polarization satisfies
\vskip 0.1cm
{\item{{\it i.}}{${\hat {\cal F}}_g(x,\bar{x})$
is invariant under $\Gamma$ --- that is, it is a modular form
of $\Gamma$ of weight zero.}}
\vskip 0.1cm
{\item{{\it  ii.}}{${\hat {\cal F}}_g(x,\bar{x})$
is ``almost'' holomorphic --- its anti-holomorphic dependence can be summarized in a finite power series in
$(\t-{\bar \tau})^{-1}$.}}
\vskip 0.1cm
On the other hand, the topological string partition function in $real$
polarization satisfies
\vskip 0.1cm
{\item{{\it iii.}}{
${\cal F}_g(x)$ is holomorphic, but not modular in the usual sense.}}
\vskip 0.25cm
{\item{{\it iv.}}{${\cal F}_g(x)$
is the constant part of the series expansion of ${\hat {\cal F}}_g(x,\bar{x})$ in $(\t-{\bar \tau})^{-1}.$}}
%
%\lref\kz{Zagier}
\vskip 0.2cm
Forms of this type were considered by Kaneko and Zagier \kz.\foot{To
be precise, \kz\ considers only modular forms of $SL(2,\IZ)$. However,
this has an obvious generalization, at least in principle, to (subgroups of) $Sp(2n,\IZ)$.}
In \kz\ forms satisfying $i.$ and $ii.$ (with arbitrary weight) are
called {\it almost holomorphic modular forms} of $\Gamma$. Moreover,
for every almost holomorphic modular form, \kz\ defines the associated
{\it quasi-modular form} as that satisfying $iii.$ and $iv.$ These are
holomorphic forms which are not modular in the usual sense.
%\vskip 0.1cm
This suggests that the genus $g$ amplitudes are in fact
naturally (almost) modular
functions of $\tau$ (and $\bar{\tau}$ in holomorphic polarization),
which can be extended from functions on the moduli space
${\cal M}$ of complex
structures to the space ${\cal H}_{X}$ parameterized by the period matrix
$\tau_{IJ}$ on $X$ modulo $\Gamma$.
In the following, we will mainly study this in the local Calabi-Yau examples, and show
that this indeed is the case, leaving compact Calabi-Yau manifolds for
future work.

Now, take a holomorphic, quasi-modular form
$E_{IJ}(\tau)$ of $\Gamma$,
such that
\eqn\prop{
{\hat E}^{IJ}(\tau,\bar \tau) = E^{IJ}(\tau)+ \left(( \tau-{\bar \tau})^{-1}\right)^{IJ}
}
is a modular form, albeit an almost holomorphic one.
Since $(\tau-\bar\tau)^{-1}$ transforms under $\Gamma$ as in
\anom , for ${\hat E}^{IJ}$ to be modular,
$E^{IJ}$ must transform as
\eqn\Etrans{
E^{IJ}(\tau) \rightarrow
{(C\t +D)^{I}}_{K}\;{(C\t+D)^{J}}_L \;E^{KL}(\tau)+
C^{IL}{(C\t+D)^{J}}_L.
}
Then $\hat{E}$ transforms simply as
\eqn\antihol{
{\hat E}^{IJ}(\tau, \bar{\tau}) \rightarrow
{(C\t +D)^{I}}_{K}\;{(C\t+D)^{J}}_L \;{\hat E}^{KL}(\tau,\bar{\tau})
}
Of course, $E^{IJ}$ and ${\hat E}^{IJ}$ are just $\Gamma \subset
Sp(2n, \IZ)$
analogues (up to normalization) of the second Eisenstein series $E_2(\tau)$ of $SL(2,
\IZ)$, and its modular but non-holomorphic counterpart $E_2^*(\tau,{\bar \tau})$ --- see Appendix A.
It is important to
note that the transformation properties given above $do$ $not$ define $E$ and $\hat{E}$
uniquely: shifting $E^{IJ}$ by any holomorphic modular form $e^{IJ}$
of $\Gamma$,
$$
E^{IJ}(\tau) \rightarrow E^{IJ}(\tau) + e^{IJ}(\tau)
$$
with $e^{IJ}(\tau)$ transforming as
$$
e^{IJ}(\tau) \rightarrow
{(C\t +D)^{I}}_{K}\;{(C\t+D)^{J}}_L \;e^{KL}(\tau),
$$
we still get a solution of \Etrans.

With this in hand, one can reorganize each ${\cal F}_g$ as a finite
power series in ${E}$ with coefficients that are strictly
holomorphic modular forms \kz . In particular, the free energy at
genus $g$ in holomorphic polarization can be written as
\vskip 0.01cm
\eqn\modular{
{\hat{\cal F}}_g(\tau,{\bar \tau}) = h_g^{(0)}(\tau) +
({h_g^{(1)}})_{IJ}\, \hat{E}^{IJ}(\tau,\bar{\tau})+\ldots
+ ({h_g^{(3g-3)}})_{I_1\ldots I_{6g-6}}\,
\hat{E}^{I_1I_2}(\tau,\bar{\tau})\ldots
\hat{E}^{I_{6g-7}I_{6g-6}}(\tau,\bar{\tau}),
}
%\vskip 0.01cm
%
where $h_g^{(k)}(\tau)$ are holomorphic modular forms of $\Gamma$ in
the usual sense.
Moreover,
taking ${\hat{\cal F}}_g(\tau,{\bar \tau})$
and sending $\bar \tau$ to infinity,\foot{By sending $\bar \tau$ to infinity what we really mean is keeping the constant term in the finite power series in $(\tau-\bar \tau)^{-1}$. For $SL(2,\IZ)$, this is simply the isomorphism between the rings of almost holomorphic modular forms and quasi-modular forms described in \kz, which can be easily generalized to $Sp(2n,\IZ)$.}
$$
{\cal F}_g(\tau) = \lim_{\bar \t \rightarrow \infty}
{\hat{\cal F}}_g(\tau,{\bar \tau})
$$
we recover
the modular expansion of the partition function in real polarization:
\vskip 0.1cm
$$
{{\cal F}}_g(\tau) = {h}_g^{(0)}(\tau) +
({h_g^{(1)}})_{IJ}\, {E}^{IJ}(\tau)+\ldots
+ ({h_g^{(3g-3)}})_{I_1\ldots I_{6g-6}}\,
{E}^{I_1I_2}(\tau)\ldots
{E}^{I_{6g-7}I_{6g-6}}(\tau).
$$
\vskip 0.1cm
\vskip 0.1cm
This gives us a way to construct modular invariant quantities out of
the free energy and correlation functions.
For example, it is easy to see that the highest order term
in the $(\tau-\bar\tau)^{-1}$ expansion of
${\hat \F}_g$ is always modular.
It is constructed solely out of genus zero amplitudes, as it
corresponds to the most degenerate genus $g$ Riemann surface that
breaks up into $(2g-2)$ genus zero components with three punctures
each.  Moreover, it follows that ${\del_I\del_J\del_K} {\F}_0$
is itself modular and corresponds to an irreducible
representation --- a third rank symmetric tensor:
\eqn\Cthree{
{\del_I\del_J\del_K} {\F}_0
\rightarrow
\left(( C\t +D)^{-1}\right)^{I'}_{\;\;I}\;
\left((C\t+D)^{-1}\right)^{J'}_{\;\;J} \;
\left((C\t+D)^{-1}\right)^{K'}_{\;\;K} \;
{\del_{I'}\del_{J'}\del_{K'}} {\F}_0,
}
which can be verified directly as well.
\vskip 0.1cm

{}From $h_g^{(0)}$, we get a modular forms of weight zero, constructed out of ${\cal F}_g$ and lower genus
amplitudes via
\eqn\agtwo{
(h_g^{(0)})(\tau)=  {\cal F}_g(\tau) +\Gamma_g({E}^{IJ}(\tau),~\del_{I_1}\ldots\del_{I_N} {\cal F}_{r<g}),}
where $\Gamma_g$ is the functional introduced in the previous sections.
While none of the terms on the right hand side is modular
on its own, added together we get a modular invariant of $\Gamma$.
We can turn this around and read this equation as follows: given the genus $r < g$ amplitudes and the propagator
$E^{IJ}$,
the free energy ${\cal F}_g(\tau)$
is fixed, up to the addition of a precisely
modular holomorphic form $h_g^{(0)}$! 
In practice, this means that 
$h_g^{(0)}$ is a meromorphic function on the moduli space.\foot{As stated in section 2, throughout we assumed a definite choice of a gauge, and picked a 3-form $\omega$ as a definite section of ${\cal L}$. Like ${{\cal F}_g}$'s, $h_g^{(0)}$ depend on this choice -- they 
are sections of 
${\cal L}^{2-2g}$, so $h_g^{(0)}$ is 
more precisely a meromorphic section of ${\cal L}^{2-2g}$. On a non-compact 
Calabi-Yau, however, it is simply a meromorphic function.}

We can write this compactly as follows.
Let
$$
{\cal H}(\tau)
= \sum_{g=1}^{\infty}\, h^{(0)}_g (\tau)\;g_s^{2g-2}
$$
be the generating functional of weight zero modular forms, and
define the generating function of correlation functions
$$
{\cal W}(y, x) = \sum_{g,n}\; {1\over n!}
\; \del_{I_1}\ldots \del_{I_n} {\cal F}_g(x)  \;
y^{I_1}\ldots y^{I_n}\;g_s^{2g-2}
$$
where the sum over $n$ runs from zero to infinity, except at genus
zero and one, where it starts at $n=3$ and $n=1$, respectively.
Then, the above can be summarized by writing
$$
\exp(\,{\cal H}(x)) = \int \;dy \;\exp(\,-{1\over 2 g_s^2}\,E_{IJ} \,y^I y^J)\; \exp(\,{\cal W}(y,x)\,)
$$
where $E_{IJ}$ is the inverse of $E^{IJ}$,
$$
E^{IK} E_{KJ} = \delta^{I}_{J}.
$$
This follows directly from the path integral of section 2
relating the wave functions in the real and holomorphic polarizations,
which we can be written as
$$
\hat{Z}(x,\bar x) = \int dy (-\,{1\over 2 g_s^2}\,(E-{\hat E})_{IJ} \,y^I y^J)\; \exp(\,{\cal W}(y,x)\,)
$$
where one views ${\hat E}$ as a perturbation.

Furthermore, one can show that similar equations
hold when ${\cal F}$ and $E$ are replaced
by their non-holomorphic counterparts.
To see this, note that the inverse of \delk\ is
\eqn\delkinv{
Z(x)= \int\, dz\; e^{{\hat S}(x,z)/g_s^2} \,
{\hat Z}(z;X,{\bar X} ),
}
with all the quantities as defined in section 2.
If we choose the background $X^I=x^I$, this
has a saddle point at $z^{I} = x^I.$
%where $z^I=\varphi x^I + z^{i}D_i x^I$.
Expanding around it, by putting $z^{I} = x^I + y^I$
where $y^I =-\varphi x^I + z^{i} D_i x^I$,
and integrating over $y$, we get
$$
Z(x)= \int\, dy\;
\exp(-{1\over g_s^2} \,(\tau - \bar{\tau})_{IJ}y^{I} y^I)
\exp(\,{\hat{\cal  W}}(y;x,{\bar x} )\,),
$$
where
$$\eqalign{
{\hat{\cal  W}}(y; x,{\bar x})& = \,\sum_{g}\; \;g_s^{2g-2}\;
{\hat{\cal F}}_g((1-\varphi)x+ z^iD_i x,\bar{x})\cr
&=
\sum_{n,g}\; {1\over n!}\, g_s^{2 g-2}\;(1-\varphi)^{2-2g-n}\;
z^{i_1} \ldots z^{i_n}D_{i_1} \ldots
D_{i_n} \hat{ {\cal F}}_g(x,\bar{x}) -
({\chi\over 24} -1) \log(1-\varphi).
}
$$
{}From this, and thinking about $Z(x)$ in terms of a power series in $E$, it follows immediately that
\eqn\bcoveq{
\exp({\cal H}(x)) = \int \;dy \;\exp(\,-{1\over 2 g_s^2}\,{\hat E}_{IJ}(x,\bar x) \,y^I y^J)\; \exp(\,{\hat {\cal W}}(y,x,{\bar x})\,).
}
The equation \bcoveq\ has appeared before. In the seminal paper \BCOV\
the authors derived a set of equations that the physical free energies
${\hat {\cal F}}_g$ must satisfy, through analysis of the
worldsheet theory. These equations were interpreted in \wittenBI\ as
saying that the topological string partition function is a wave
function in the Hilbert space obtained from the geometric quantization
of $H^3(X,{\IC})$, the fact that we used repeatedly here.
%But, \BCOV\
%also proposed a way to solve the holomorphic anomaly equations, which,
%almost miraculously, fixed the ${\hat{\cal F}}_g$'s, up to a
%holomorphic function on the moduli space. The authors stated this
%compactly in equation $(6.14)$ of that paper, which is easily
%recognized in \bcoveq\ --- however,  we derived it in a somewhat
%more transparent way, by simple symmetry considerations.
Holomorphic anomaly equations (and modular invariance) constrain what the topological string amplitudes can be. Here we described the solutions to the equations
using symmetry alone.
The construction of the propagators ${\hat E}$,
which was the guts of the method of \BCOV\ for solving the equations,
was quite complicated. The answers were messy,
with ambiguities that had no clear interpretation.
%It is fair to say that this was a
%not so appealing aspect of an otherwise beautiful work.
Now, the meaning of the propagators $\hat{E}^{IJ}$ and
$E^{IJ}$ is simple and beautiful --- they are simply generators of (almost) modular
forms of the symmetry group $\Gamma$!

The only remaining thing to show is that the
propagators of our expansion and of \BCOV\ agree. In \BCOV\ the
authors gave a set of relations that the inverse propagators satisfy
(p. 103 of \BCOV ). It is easily shown that our propagators
\prop\ satisfy
these relations (for any holomorphic form $E_{IJ}$). Let
$$
{\hat E}_{\vphi\vphi} = {\hat E}_{IJ}\, x^{I}\, x^{J},
\qquad
{\hat E}_{\vphi i} = {\hat E}_{IJ}\, x^{I}\, D_i x^{J},
\qquad
{\hat E}_{ij} = {\hat E}_{IJ}\, D_i x^{I}\, D_j x^{J},
$$
where $D_i$ is the K\"ahler covariant derivative
$D_i = \del_i - \del_i K$
and $K$ is the K\"ahler form of the special geometry of $X$.
Then, with a bit of algebra it follows that
these satisfy
\eqn\ours{
\eqalign{
{\bar \del}_{\bar i}
{\hat E}_{j k}& =
{\bar C}_{\bar i}^{mn}
{\hat E}_{m j}{\hat E}_{n k} +
{G}_{\bar i j} {\hat E}_{\vphi k}+
{G}_{\bar i k} {\hat E}_{\vphi j}\cr
{\bar \del}_{\bar i}
{\hat E}_{j \varphi}& =
{\bar C}_{\bar i}^{mn} {\hat E}_{m j}{\hat E}_{n \vphi} +
{G}_{\bar i j} {\hat E}_{\vphi \vphi}\cr
{\bar \del}_{\bar i}
{\hat E}_{\vphi \vphi} &=
{\bar C}_{\bar i}^{mn} {\hat E}_{m \vphi}{\hat E}_{n \vphi}
}
}
where
$$
G_{{\bar i}j} = {\bar \del}_{\bar i} \del_j K,
\;\; {\bar C}_{\bar i}^{mn} =
e^{-2 K} G^{m{\bar m}} G^{n{\bar n}}
{\bar C}_{\bar{i}\bar{m}\bar{n}},
\;\;
{\bar C}_{\bar{i}\bar{m}\bar{n}} =
{\bar C}_{IJK}
{\bar D}_{\bar i} {\bar x}^{I}
{\bar D}_{\bar j} {\bar x}^{J}
{\bar D}_{\bar k} {\bar x}^{K}.
$$
%
%{\bar C}_{IJK} =
%\bar{\del_I} \bar{\del_J} \bar{\del_K} \bar{{\cal F}}_0
%
The equations \ours\
are exactly the equations of \BCOV\ with obvious substitutions.

\subsec{A Mathematical Subtlety}

As we have shown in the previous sections, our results are
completely general and apply to both non-compact and compact
Calabi-Yau threefolds. However, to make contact with the theory of
modular forms in mathematics there is an important subtlety that we
have not mentioned yet.

In the theory of modular forms, the period matrix $\tau_{IJ}$ acquires a crucial role. A modular form is defined to be a holomorphic function $f: {\cal H}_k \rightarrow \IC$ satisfying certain transformation properties, where ${\cal H}_k$ is the Siegel upper half-space:
$$
{\cal H}_k = \{\tau \in {\rm Mat}_{k\times k} (\,\IC)
| \,\tau^T = \tau, \tau - \bar \tau > 0\},
$$
which is the space of $k \times k$ symmetric matrices with positive definite imaginary part. The period matrix is the $\tau$ in the definition of the Siegel upper half-space. Note that strictly speaking, this defines Siegel modular forms; proper modular forms are obtained for $k=1$.\foot{See Appendix A and B for definitions and conventions.}

For the non-compact case, the mirror symmetric geometry reduces to a family of Riemann surfaces of a certain genus. Thus, it is clear that the period matrix $\tau_{IJ}$ has positive definite imaginary part. Therefore, in this case our results should be interpreted mathematically as Siegel modular forms, where $k$ depends on the genus $g$ of the Riemann surface. In particular, if the mirror geometry is a family of elliptic curves, $k=1$, and we recover proper modular forms.

However, in the compact case the situation changes slightly. The
period matrix $\tau_{IJ}$ does not have positive definite imaginary
part anymore; it has signature $(h^{2,1},1)$, as explained for
instance in \DVV. Thus, in this case the Siegel upper half-space is
not the relevant object anymore, and we cannot make contact directly
with Siegel modular forms. This seems to call for a new theory of
modular forms defined on spaces with indefinite signature. It would be
very interesting to develop this mathematically.

Another possibility, in order to make
contact with already known mathematical concepts
in the compact case, is to replace the period matrix $\tau_{IJ}$ by a
different but related matrix ${\cal N}_{IJ}$ --- see for instance
\DVV\ for a definition --- which has positive definite imaginary part,
but is not holomorphic. This is usually done in the context of
supergravity. Roughly speaking, it amounts to replacing the
intersection pairing by the Hodge star pairing. In that way perhaps we
can come back into the realm of Siegel modular forms, perhaps along the lines of what was done in \DVV\ in a related context.

In the following sections we will give applications of the modular
approach we have developed so far, for local Calabi-Yau threefolds.

%%%%%%%%%%%%%%%%%%%%%%%%%%%%%%%

\newsec{Seiberg-Witten Theory}

As is well known, type II string theory compactified on
local Calabi-Yau manifolds
gives rise to ${\cal N}=2 $ gauge theories in four dimensions.
The topological string theory on these manifolds computes topological
terms in the effective action of $N=2$ Seiberg-Witten theory with
gauge group $G$ \refs{\BCOV,\KatzFH}.  These terms are summarized in a
partition function
\eqn\zsw{
Z_{SW}=\exp(\lambda^{2g-2} \F_{g}(a))\ ,
}
where $\F_g$ coincides with the genus $g$ topological string free
energy, and the $a$'s are local parameters in the vacuum manifold of
the gauge theory.  Each term in \zsw~has a physical meaning in the
effective action of the ${\cal N}=2$ gauge theory. The genus zero
topological string amplitude yields the exact gauge coupling
\eqn\tausw{\tau_{ij}=\frac{\partial^2 \F_{0}}{\partial a_i\partial a_j}\ ,}
with $i,j=1,\ldots r$, where $r= {\rm rank}(G)$, while the higher
genus topological string amplitudes yield the gravitational coupling
of the self-dual part of the curvature $R_+$ to the self-dual part
of the graviphoton field strength $\int {\rm d} x^4 \F_{g} R^2_+
F^{2g-2}_+$. The $\F_{g}(a)$'s for $g>1$ were in fact extensively
studied in the weak electric coupling limit \NekrasovRJ.

The corresponding Calabi-Yau manifold is given by an equation of the
form \LocalCY\ with an appropriate $H(y,z)$ depending on the
theory. For example, for $G=SU(n)$ without matter,
\eqn\curve{
H(y,z) = y^2 - (P_n (z))^2 + 1
}
where $P_n(z) = z^n + u_2 z^{n-2} + \ldots u_{n}$, and the holomorphic
3-form is given by \three.
The parameters $u_i$ are complex coordinates on the moduli space of the Calabi-Yau. In the gauge theory, they correspond to
the expectation values of the gauge invariant observables
\eqn\gaugeinvariantu{u_k={1\over k} {\rm Tr} \langle \phi^{k}\rangle +{\rm products\ of \
lower\ order\ Casimirs},}
where $\phi$ is the adjoint valued Higgs field.

The family of Riemann surfaces
obtained by setting
$$
\Sigma_g: \qquad H(y,z)=0
$$
is the Seiberg-Witten curve of the gauge theory.  The genus $g$ of the
Riemann surface is the rank of the gauge group $r$.  The gauge
coupling constant ${\rm Im}(\tau_{ij})$ is the period matrix of the
Riemann surface.  Alternatively, $\tau_{ij}$ is the complex structure
of the Jacobian of the Riemann surface ${\Sigma}_g$, which is an
abelian variety.  The abelian variety is spanned by the periods
\eqn\periods{\pmatrix{a_{D_i}\cr a_i}=\pmatrix{p_i\cr x^i}=\pmatrix{
\int_{B_i} \lambda\cr \int_{A^i} \lambda}\ ,
}
with $i=1,\ldots r$, and where the A- and B-cycles generate
the symplectic integer basis of $H_1({\Sigma}_g,\IZ)$.  Here $\lambda$
is a meromorphic differential, which is part of the data of the
theory. As explained in section 2, in the string theory context,
$\lambda$ comes from the reduction of the holomorphic 3-form of the parent
Calabi-Yau threefold to a one-form on $\Sigma_g$.
For theories with matter, there can be additional periods on $\Sigma_g$ ---  $\lambda$ then  has poles
whose residues correspond to the mass
parameters.

The monodromy group $\Gamma$ of the curve $\Sigma_g$, which is naturally a subgroup of ${\rm Sp}(2 r,\IZ)$, played
the central role in \refs{\SW}.  It is generated by the BPS particles going massless at a codimension one loci in the moduli space and
captures the non-perturbative duality symmetries of the ${\cal
N}=2$ gauge theory, since it acts non-trivially on the coupling constant
$\tau_{ij}$.  From the monodromies of the periods around the
perturbative limits in the moduli space, \SW\ showed that one can
deduce the periods themselves everywhere in the moduli space --- this
is the Riemann-Hilbert problem --- and hence also ${\tau_{ij}}$ and
${\cal F}_0$. It is then very natural to ask what does the group
$\Gamma$ of symmetries imply about the full partition function
$Z_{SW}$.  In fact, this question, and the close relation of Seiberg-Witten theory and topological strings in general, is what motivated this
paper.

The topological string partition function is a wave function for
both compact Calabi-Yau threefolds, studied in \BCOV, and
non-compact Calabi-Yau threefolds, as we have seen in section 2.
This implies that the Seiberg-Witten partition function
\wittenBI~$Z_{SW}$ is a wave function, arising by geometric
quantization of $H_1({\Sigma}_g)$ --- see \HK. In particular, in
holomorphic polarization, it satisfies the local holomorphic anomaly
equation \lhe . In fact, it would be very interesting to derive this
directly from the ${\cal N}=2$ gauge theory.

Since the partition function $Z_{SW}$ is known, this
gives us a testing ground for exploring the restrictions that
follow from the duality symmetries generated by $\Gamma$, but now
acting on the full quantum wave function $Z_{SW}$.\foot{The observation that duality transformations imply quasi-modular properties of the $\F_g$'s has been made earlier in \deWit. However, their results are different from ours in that their partition function $Z=\exp \F$ does not transform like a wave function; rather, it transforms by Legendre transformations of $\F$.}

\subsec{Seiberg-Witten Theory and Modular Forms}

One crucial property
of the abelian variety is that ${\rm Im}(\tau_{ij})>0$, which
ensures positivity of the kinetic terms of the vector multiplet.
Thus, in this case the period matrix $\tau_{ij}$ can be used to define
the Siegel upper half space
${\cal H}_r$ as
\eqn\siegelupper{
{\cal H}_r = \{\tau \in {\rm Mat}_{r\times r} (\;\IC) | \tau^T = \tau, {\rm Im}(\tau) > 0\}.}
The monodromy group $\Gamma\subset Sp(2r,\IZ)$ of the family of
Riemann surfaces $\Sigma_g$ acts on $\tau_{ij}$ as
$$
\tau \rightarrow (A \tau + B)(C \tau + D)^{-1}~~~~{\rm for}~~~ \pmatrix{A & B \cr C & D} \in \Gamma.
$$
Thus, in principle, we should be able to give explicit expressions for
the Seiberg-Witten higher genus amplitudes in terms of Siegel modular
forms under the corresponding subgroup $\Gamma\subset Sp(2r,\IZ)$
(see appendix B for a brief review of Siegel modular forms). To start with, however, let us
consider $SU(2)$ gauge theory, where the modular group $\Gamma \subset
SL(2,\IZ)$, and correspondingly standard modular forms suffice.

\vskip 0.3cm
\item{{\it i}.}{$SU(2)$ Seiberg-Witten theory}
\vskip 0.3cm

The curve of the $SU(2)$ gauge theory can be written as\foot{As
explained in \refs{\SWtwo} there are two curves corresponding to this
gauge theory, differing by a factor of 2 in the normalization of the
$A$-period and electric charge. The curve at hand has $\#(A \cap B) =
2$ between the generators of $H_1(\Sigma,\IZ)$. The curve which is the
$n=2$ specialization of \curve\ has the A-period $A' =
A/2$. Correspondingly, the modular groups will differ: in the second
case we would get the $\Gamma_0(4)$ subgroup of $SL(2,\IZ)$ instead of $\Gamma(2)$.}

\eqn\curvestwo{
y^2 = (x^2-1)(x-u).
}
There are three singular points in the moduli space, corresponding to
$u=\pm 1, \infty$ with monodromies
\eqn\monSW{
M_\infty = \pmatrix{ -1& 2\cr
0 &-1}, \qquad
M_1 = \pmatrix{ 1& 0\cr
-2  &1}, \qquad
M_{-1} = \pmatrix{ -1& 2\cr
-2 &3}
}
acting on
$$
\Pi = \pmatrix{p \cr x} = \pmatrix{\int_B \lambda \cr
\int_A \lambda}
$$
where
\eqn\intersSW{
\# (A \cap B) = 2.
}
The monodromies \monSW\ generate the $\Gamma(2)$ subgroup of $SL(2,\IZ)$; that is, the subgroup of $2 \times 2$ matrices congruent to the identity matrix, modulo $2$.
The $x= a$, $p= a_D$ are by now canonical variables of Seiberg-Witten theory
\refs{\SW},  so we will mainly use that notation.

The periods $a$, $a_D$ solve the Picard-Fuchs equation
$$
{\cal L} \Pi =0,
$$
where ${\cal L} = \theta (\theta - 1) - u^2(\theta -{1\over 2})^2$
and $\theta = u {\del \over \del u}$.
{}From the previous sections, we can predict that the genus $g$
amplitudes $\F_g$ of this theory are (almost) modular forms of
$\Gamma(2)$, with definite transformation properties.
Since the higher genus amplitudes are
known from \refs{\SW, \nekrasov}, they will
provide a direct check of our predictions.

The parameter $\tau$ of the modular curve is defined by
$\tau = {\del p \over \del x}$, or in usual Seiberg-Witten notation
\eqn\tauswtwo{
\tau = {\del a_D \over \del a} = {2} {\del^2 \over \del a^2} \F_0(a).
}
Solving the Picard-Fuchs equation for the periods, we can obtain
$\tau$ as a function of $u$.  Alternatively, we can proceed as
follows. Recall that the $j$-function of the elliptic curve, which has the $q$-expansion
$$
j(\tau) = {1\over q} + 744 + 196884 q+ \ldots
$$
where $q = e^{2 \pi i \tau}$,
provides a coordinate independent way of characterizing the
curve. Roughly speaking, elliptic curves are the same if their $j$-functions are equal.
Bringing the equation \curvestwo\
of the family of elliptic curves in Weierstrass form
\eqn\Weierstrass{
y^2 = 4 x^3 - g_2 x - g_3
}
the $j$ function can be computed as
\eqn\jfn{
j = 1728 {g_2^3 \over g_2^3 - 27 g_3^2}.
}
For the family of elliptic curves \curvestwo, this gives
\eqn\jsw{
j(\tau) = { 64 (3+u^2)^3 \over (u^2-1)^2}.
}
Then, using the $q$-expansion of the $j$-function, we get a $q$-expansion for $u$, in the large $u$ limit:
$$
u = {1 \over 8}q^{-1/2} +{5 \over 2}q^{1/2} -{31 \over 4} q^{3/2} + 27 q^{5/2} +  {\cal O}(q^{7/2}).
$$

However, what we want is an expression of $u$ in terms of $\tau$ which is
valid everywhere in the moduli space, not just a $q$-expansion when $u$ is large; in other words, we want to find the modular form of
$\Gamma(2)$ which has the above $q$-expansion.
Since $u$ is a good coordinate on the moduli space,
which is the quotient of the Teichmuller space by $\Gamma(2)$,
it has to be invariant under $\Gamma(2)$; i.e., it must be a
modular form of weight zero. For a brief review of modular forms of $\Gamma(2)$, see Appendix A.

The modular forms of $\Gamma(2)$ are generated by the following $\theta$-constants:
$$
b(\tau):=\theta_2^4(\tau),\qquad
c(\tau):=\theta_3^4(\tau),\qquad
d(\tau):=\theta_4^4(\tau)
$$
which all have weight $2$. These are not independent, but satisfy the relation
$$
c= b+d.
$$
It is easy to show that \HK\
\eqn\utau{
u(\tau) ={c+d\over b} (\tau),
}
which is modular invariant, as claimed.

The genus one amplitude \refs{\MW}
\eqn\holFI{{\cal F}_{1}=-{1\over 2}\log\left(\det\left(\partial a\over
\partial u\right)\right)-{1\over 12} \log(u^2-1)\
}
can be rewritten, using the results we have obtained so far,
as \refs{\klemmmarinotheisen}
\vskip 0.2cm
\eqn\foneSW{
\F_1(\tau) = - \log \eta(\tau)
}
where $\eta(\tau)$ is the Dedekind $\eta$-function. Note that this transforms under modular transformation in $\Gamma(2)$
exactly as predicted in section 2, namely
$$
\F_1\left({A \tau + B \over C \tau +D} \right)
= \F_1(\tau) + {1\over 2} \log{{1 \over \tau + C^{-1}D}}
$$
(up to a constant that is irrelevant, as only $\del \F_1$ is well defined).\foot{In this case, $\F_1$ transforms in this way under the whole
$SL(2,\IZ)$, but this is an accident of the model. In particular, had we worked with $\Gamma_0(4)$ (and hence with $\tau'=\tau/2$),
$\F_1$ would transform like this under $\Gamma_0(4)$, but not under the full $SL(2,\IZ)$.}

Next, from section 4, we expect that ${\del^3 {\F}_0 \over \del a^3} =
{1 \over 2} {\del \tau \over \del a}$ is a modular form of weight
$-3$. Using the fact that ${\del \over \del a} = {\del u \over \del
a}{\del \over \del u}$, the modular expression for $u$ \utau\ and the
modular expression for ${\del u \over \del a}$ obtained by combining
\holFI\ and \foneSW, we get
\eqn\fzeroSW{
{\del^3 \over \del a^3} {\F}_0 (a)=  - {\sqrt{b}\over c\; d}
}
which indeed transforms as expected.
%%%%%%%%%%%%%%%%%%%%%%%%%%%%%%%%%%%%%%%%

Now, consider the genus two amplitude. In
\HK\ it was
shown that this can be written
as
\eqn\ftwoSW{
{\F}_2(\tau)=
h_2^{(0)}(\tau)+  h_2^{(1)}(\tau) \;E(\tau) +
h_2^{(2)}(\tau)\; (E(\tau))^2+  h_2^{(3)}(\tau)\; (E(\tau))^3
}
\vskip 0.2cm
\noindent where the propagator $E(\tau)$ is given in terms of the second Eisenstein series
$$
E(\tau) = {2 \pi i \over  6} E_2(\tau),
$$
and the modular coefficients are
\eqn\hs{\eqalign{
h_2^{(0)} &=  {1\over 30} (c+d) (16 b^2 +19 c d )\, X\cr
h_2^{(1)} &= - 2 \, \left({6 \over 2 \pi i}\right)\,   (b^2 +c d )\,X\cr
h_2^{(2)} &= 3 \, \left({6 \over 2 \pi i}\right)^2  \, (c +d)\,X \cr
h_2^{(3)} &= - {5\over 3}\, \left({6 \over 2 \pi i}\right)^3 \,X
}}
where we defined
$$
X = {1\over 1728} {b\over c^2 d^2}.
$$
%\alpha:=\exp\left({2 \pi i\over 3}\right)\ .
%$$
%These $\theta$-functions generate modular forms on $\Gamma(3)$.
%
We will now see that this is exactly as predicted in section 4!

First, consider how this transforms under modular transformations in $\Gamma$. Note that the coefficients $h_2^{(k)}$
are modular forms of $\Gamma$ of weight $(-3 k)$:
$$
h_2^{(k)}( \,(A \tau +B)/(C\tau+D)\, ) = (C \tau + D)^{-3 k}
\, h_2^{(k)}(\tau)
$$
Moreover, $k$ ranges from zero to $3g-3$, where $g=2$ in this case.

On the other hand $E(\tau)$ transforms as a quasi-modular form:
\eqn\propSW{
E((A \tau +B)/(C\tau+D)) = (C \tau + D)^2 E(\tau) +2\, C (C\tau+D);
}
in other words it is a holomorphic form, modular up to shifts
(cf. \Etrans ). The fact that ${\cal F}_2$ is a finite power series in
$E(\tau)$, with coefficients that are strictly modular forms of
$\Gamma(2)$ means that ${\cal F}_2$ is itself a quasi-modular
form of $\Gamma(2)$, per definition.  Note that the propagator in
\propSW\ transforms by a factor of 2 relative to \Etrans . This factor
of two is a consequence of the fact that the
intersection number of the A and the B periods of the curve is
twice bigger than the conventional one
\intersSW . It is very easy to derive this from section 2 and 3
(see footnote 6).

%%%%%%%%%%%%%%%%%%%%

Moreover, it is easy to check, starting from \foneSW, \fzeroSW\  and \ftwoSW\
(with the help of some standard modular formulae given in appendix A),
that $\F_2$ transforms under modular transformations
exactly as predicted in section 3. To do so, note that, looping around
$u=1$ for example, simply acts on $\tau$ by the $\Gamma(2) \subset
SL(2,\IZ)$ transformation $M_1$ given in \monSW. Using the usual transformation properties of modular forms and the expression \ftwoSW\ for $\F_2$ in terms of modular forms of $\Gamma(2)$, it is then easy to work out the transformation property of $\F_2$ under $M_1$.

Furthermore, while the $\F_g$ and the vertices $\del_{i_1},\ldots,\del_{in} \F_g$
are not quite modular, the combinations
\eqn\agtwoa{
{\cal F}_g(\tau) +\Gamma_g({E}(\tau),~\del_{i_1}\ldots\del_{i_n} {\cal F}_{r<g}) \; = \; h_g^{(0)}(\tau)}
are exactly invariant under modular transformations and agree with $h_g^{(0)}(\tau)$,
as expected from section $4$.
%Finally, note that
%
%$$
%E(\tau)=  -4 \del_{\tau} \F_1(\tau),
%$$
%
%
%as expected from section 4. The factor of $2$ relative to \proptwo\ is
%again a consequence of the unconventional intersection
%numbers.\foot{Another way to derive \proptwo\ which makes this easier
%to see is to note that the holomorphic, modular invariant quantity
%built from $\del_I {\cal F}_1$ is $\del_I {\cal F}_1 + {1\over 2} E^{JK}
%\del_{I}\del_{J}\del_{K} \F_0$, for some $E^{JK}$, defined by \Etrans
%. We can pick a $E^{IJ}$ such that the modular ambiguity
%vanishes. Now since $\del_{I}\del_J\del_K \F_0 = \del_I \tau_{JK}$ with
%the conventional intersection numbers, it yields $E_{IJ} = - 2 {\del\over
%\del \tau_{IJ}} \F_1$; this is just \proptwo . However, in the
%present case, $\del^3_{a} \F_0 = {1 \over 2} \del_a \tau$
%since $\tau = 2{\del^2\over \del a^2}\F_0(a)$, and so we must take $E=- 4 {\del\over
%\del \tau} \F_1$.}

We can trade quasi-modular forms for almost holomorphic
forms by
replacing $E(\tau)$ in all formulae
by its modular, but not holomorphic counterpart
$$
\hat{E}(\tau,\bar{\tau}) = E(\tau) + {2\over \tau - \bar\tau}
$$
which transforms as
$$
{\hat E}(\;
(A \tau +B)/(C\tau+D),\;(A {\bar \tau} +B)/(C{\bar \tau}+D)\;)
= (C \tau + D)^2 {\hat E}(\tau,\bar\tau).
$$
Also, note that ${\F}_1$ can be made exactly modular by
writing
$$
\hat{\F_1}(\tau,\bar{\tau}) = -
\log((\tau - \bar\tau)^{1\over 2}|\eta(\tau))|^2) .
$$
This is exactly the one-loop amplitude of
the local Calabi-Yau in holomorphic polarization.
More precisely, it is only the holomorphic derivatives
${\del\over \del a} \F_1$, and ${\del\over \del a} {\hat \F}_1$
that are physical, but this is the natural way
to write it.

Finally,
$
{\hat E}(\tau,\bar{\tau})
$
is exactly the propagator of \BCOV ! One has that
\eqn\agtwob{
{\hat {\cal F}}_g(\tau,\bar \tau) +
\Gamma_g(\hat{E}(\tau,\bar \tau),~\del_{i_1},\ldots,\del_{i_n} {\cal F}_{r<g}(\tau,\bar\tau))=h_g^{(0)}(\tau)
}
is strictly holomorphic, with the same modular form $h_g^{(0)}(\tau)$ as in \agtwoa .

In the next subsection, we
consider gauge groups of higher rank, corresponding to Riemann surfaces of genus higher than one.

\subsec{The $SU(n)$, $n>2$ Seiberg-Witten Theory}

As mentioned earlier, the Riemann surface corresponding to
$SU(n)$ Seiberg Witten theory is a genus $g=n-1$ curve
\eqn\curve{
y^2 - (P_n (z))^2 + \Lambda^{2n}=0
}
where
$$
P_n(z) = z^n + u_2 z^{n-2} + \ldots u_{n+1}.
$$
The singular loci in the moduli space correspond to
the zeroes of the discriminant
\eqn\discri{
\Delta = \prod_{i < j } (e_i(u) - e_j(u))^2
}
where $e_i(u)$ are roots of $P_n(z,u)^2 -\Lambda^{2n}$. That is, at the values of the moduli $u$ for which any pair
of roots come together $e_i(u) \rightarrow e_j(u)$, the curve becomes singular.  There is a natural basis of
$(n-1)$ $A$-cycles corresponding to pairs of branch points that pair up
as $\Lambda$ goes to zero.  This corresponds to points where the
non-abelian gauge bosons become massless in the classical theory. The
monodromy group $\Gamma \subset Sp(2g,\IZ)$ of the quantum theory can
be determined \refs{\klemmlerchetesisn},
by following the
exchange paths of the branch points.

We will leave the detailed analysis of this and the corresponding
implications for the structure of the topological string amplitudes as
an interesting exercise, and only consider briefly the one-loop
amplitude.

On general grounds
\refs{\BCOVI,\BCOV}, the one-loop amplitude in the topological
string theory has the universal form
\eqn\oneloophol{{{\cal F}}_1(\tau) = - {1\over 2}
\log(\det(D_i X))-\, {1\over 12}
\log(\Delta).
}
This result was
also derived in a purely gauge theory context in
\refs{\MW,\MarinoMoore}.
There, the authors computed the one-loop amplitude of the
(twisted) ${\cal N}=2$ gauge theory on a curved four-manifold, namely
the coefficients of the $\int R^2$ term in the effective
action. Restricting the curvature to be anti-self dual, $R_- =0$, this
is precisely the term that the topological string
computes.\foot{Practically, in terms of \refs{\MW,\MarinoMoore} this
corresponds to setting the signature $\sigma$ of the four-manifold
equal to $\sigma = - {2\over 3 } \chi$ where $\chi$ is its Euler
character. One way to see this is that it holds exactly for the
$K3$, for example, where the curvature is anti-self dual.} This gives
\eqn\holFI{{\cal F}_{1}(\tau)=-{1\over 2}\log\left(\det\left(\partial a_i\over
\partial u_k\right)\right)-{1\over 12} \log(\Delta),
}
where $\Delta$ is the discriminant of the Seiberg-Witten curve.
For example, for $G=SU(n)$ with curve given by \curve, $\Delta$ is \discri .

Note that the $u$'s are
necessarily modular invariants of $\Gamma$, as they are just
parameters entering into the algebraic definition of the curve, and
hence they do not `talk' to its periods. On the other hand,
$\Delta$  is simply a rational function of $u$, so also necessarily
a Siegel modular form of $\Gamma$ of weight zero.

To write the full amplitude in terms of modular forms,
note that from \refs{\MW,\MarinoMoore}  we have
\eqn\FItheta{\left( \det \left( \partial a_j\over \partial u_i \right)\right)^{{1\over 2}}
\Delta^{{1\over 8}}=\theta\left[0\atop \vec \delta \right](0,\tau)
}
where $\vec \delta=\left[{1\over 2},\ldots ,{1\over 2}\right]$ and we defined
the `generalized' $\theta$-functions with characteristic in appendix $B$.
As a consequence we can write
$$
{\cal F}_{1}(\tau)=-
\log\left(
{\theta \left[0 \atop \vec{\delta}\right]}
(0,\tau)\right)
+{1\over 24} \log(\Delta)\ .
$$
This is consistent
with the transformation properties of ${\F}_1$, since ${\theta \left[0
\atop \vec{\delta}\right]}$ is a scalar Siegel modular form of weight $1/2$.

%%%%%%%%%%%%%%%%%%%%%%%%%%%%%%%%

\newsec{Local $\IP^2$}

We now study the local ${\IP}^2$, from the mirror B-model point of
view.  In this case the mirror is a family of elliptic curves
$\Sigma$ with monodromy group $\Gamma(3)$. The Gromov-Witten theory
of the local $\IP^2$ at large radius was solved in
\refs{\TV,\Allloop} . Using those results, we can show explicitly
that the predictions for modular properties of the topological
string amplitudes are satisfied.

Another interesting
point in the moduli space of the local ${\IP}^2$ is the
$\;{\IC}^3/{\IZ}_3$ orbifold point. One can in principle formulate the
Gromov-Witten theory of the orbifold point as well, however the
amplitudes are not yet available \refs{\yongbin, \bryan}.  We now have
a simple prescription to carry over the large radius results to other
points in the moduli space, the orbifold point in particular, so we
can make new predictions there.

\subsec{Mirror Family of Elliptic Curves}

The mirror data is a family of elliptic curves $\Sigma$,
given by the equation
\eqn\cubic{
\sum_{i=1}^3 x_i^3 - 3 \psi \prod_{i=1}^3 x_i=0
}
in ${\IP^3}$, and a meromorphic 1-form
$\lambda = \log(x_2/x_3) dx_1/x_1$.
This has an obvious ${\IZ_3}$ symmetry
$$
\psi \rightarrow  \alpha \psi, \qquad \alpha = e^{2 \pi i/3},
$$
 since it can be undone by a coordinate transformation $x_1 \rightarrow \alpha^{-1} x_1$ that affects neither $\Sigma$ nor $\lambda$.
The discriminant
$\Delta$ of the curve is
$$
\Delta=(1-\psi^3).
$$
This vanishes at the three singular points
$\psi^3=1$,
corresponding to conifold singularities.

To make contact with standard
elliptic functions and their modular properties we make a
$PGL(3,\IC)$ transform to bring the equation of the curve
to its Weierstrass form
$$
y^2=4 x^3-g_2 x - g_3
$$
with
$$
g_2={\alpha ( 8+\psi^3)\over 2^{(2/3)} 24 \psi^3},~~~~ g_3= {8+20 \psi ^3- \psi^6\over 864 \psi^6},
$$
so that its $j$-function is given by
\eqn\j{
j(\tau)=- {27 \psi^3 (8+\psi^3)^3\over (1- \psi^3)^3}\ .
}
As usual,
\eqn\tp{
\tau= {\partial p\over \partial x}
}
is
the standard complex structure modulus of the family of elliptic curves,
where we view $\Sigma$ as a quotient of a complex plane by a lattice
generated by $1$ and $\tau$.
Here\foot{
We use $x$ to denote both the coordinate on the Riemann surface and the period of $\lambda$. It should be clear from the context which
meaning we assign to $x$.}
$$
p=\int_B \lambda(\psi), \qquad x=\int_A \lambda(\psi)
$$
%where $\omega(\psi)={\dd x\over y(\psi)}$ is the standard
%holomorphic $(1,0)$ form on $C$.
where $\lambda(\psi) = \log(x) dy/y$.
Our $j$-function
is normalized to
\eqn\jexplicite{
j={1\over q}+744 + 196884 q+{\cal O}(q^2),
}
where $q=\exp(2 \pi i \tau)$.
Combining the two expressions for the $j$-function, we find a series expansion for $\psi(q)$ in the large $\psi$ limit:
\eqn\psitau{
3 \psi={1\over q^{1\over 3}}+ 5 q^{2\over 3}- 7 q^{5\over 3}+{\cal O}(q^{8\over 3})\ .
}
Alternatively, we can obtain the same expansion by first using the Picard-Fuchs equations to find the periods $x(\psi)$, $p(\psi)$, and then computing
$\tau(\psi)$ directly using the definition \tp . We will study in more details the Picard-Fuchs equation and its solutions in the next subsection. For now, we only note one
interesting aspect to this. Namely, as discussed in section 2.3, due to the non-compactness of the Calabi-Yau, it may not be possible to find a basis of periods that are normalized canonically. This occurs in the present example: the compact B period satisfies
\eqn\inter{
\#(A\cap B) = -3 .
}
One way to see this is in the mirror A-model: the compact parts of $H_4$
and $H_2$ of the manifold are generated by the ${\IP}^2$, which we take to be mirror to the B-period,  and the $\IP^1$
line inside it, mirror to the A period. In the Calabi-Yau, these do intersect, but the intersection number is $-3$. Correspondingly, if we put $x=t$,
$$
p = -3
{\del\over dt} {\cal F}_0(t),
$$
and therefore $\tau = -3 {\del^2\over dt^2} {\cal F}_0(t)$.

The above expression for $\psi(\tau)$ is valid for
${\rm Im}(\tau)\rightarrow \infty$. In the next subsection, we will show that
the local $\IP^2$ is governed by a $\Gamma(3)$ subgroup of $SL(2,Z)$.
This will allow us to give a globally well defined expression for
$\psi$ in terms of modular forms under $\Gamma(3)$.

\subsec{The Monodromy Group}

The meromorphic 1-form $\lambda$ turns out to have a non-vanishing
residue: in addition to the usual $A$ and $B$ periods --- by this we mean the periods associated to the $A$ and $B$ cycles --- of
the genus one Riemann surface, it has an additional period, which we will
call $C$. As discussed in section 2.3, the extra period does not
correspond to a modulus of the Riemann surface, but to an auxiliary parameter.
While the monodromies mix up all the periods, the monodromy action on
the extra period $C$ should be highly constrained. To derive the monodromy action on
the full period vector
$$
{\Pi}= \pmatrix{\int_B \lambda \cr \int_A \lambda \cr \int_C \lambda}
$$
we will solve the Picard-Fuchs (PF) differential
equations that $\Pi$ satisfies
\eqn\pf{
{\cal L} \Pi =0,
}
$everywhere$ in the moduli space. A certain linear combination of the solutions to equation \pf\ will have the property that its monodromies are integral, and
that gives $\Pi$.

Before doing that, note that, since the additional period C is just an auxiliary parameter, the modular properties of the topological string amplitudes should be governed by the monodromy group
of the family of elliptic curves $\Sigma$. It is well known that this is a
$\Gamma(3)$ subgroup of $SL(2,\IZ)$, when viewed as a fibration over the punctured $\psi$ plane. We will see below that this is indeed the case.

Now let us come back to the study of the full Picard-Fuchs equation. It is convenient to work in the coordinate $z$, centered at large radius:
\eqn\zphi{
z=-{1\over (3 \psi)^3} .
}
There are three special points in the $z$ plane. In addition to
the large radius point at $z=0$, there is also the conifold point,
coming from $\psi^3=1$, and the orbifold point $z=\infty$, with
${\IZ}_3$ monodromy around it. In this coordinate
the Picard-Fuchs differential operator ${\cal L}$ has the form
$$
{\cal L}=\theta_z^3+ 3 z ( 3 \theta_z+2) (3\theta_z +1) \theta\ .
$$
This has three independent solutions, one of which is a constant,
corresponding to the period of $\lambda$ around the $C-$cycle. The
corresponding new cycle $C$ encircles the residue of
$\lambda(\psi)$.

The solutions near large radius ($z=0$) can be found by the Frobenius method
from the generating function
$$
\omega(z,s):=\sum_{n=1}^\infty {z^{s+n}\over \Gamma(- 3(n+s)+1)\Gamma^3(n+s+1)},
$$
with ${\cal L} \omega(z,s)=0$. This gives 3 independent solutions,
$$
\omega_i={1\over (2 \pi i)^i}\left. {\dd^i\over \dd^i s}\omega(z,s)\right|_{s=0},
$$
i.e. $\omega_0=1$, $\omega_1= {1\over 2 \pi i}(\log(z) + \sigma_1(z))$ and
$\omega_2= {1\over (2 \pi i)^2} (\log(z)^2+2 \sigma_1\log(z) + \sigma_2(z))$,
where the first orders are $\sigma_1=-6\,z + 45\,z^2 +{\cal O}(z^3)$ and
$\sigma_2=-18\,z + \frac{423\,z^2}{2} +{\cal O}(z^3)$.

Linear combinations of these solutions will give the periods over cycles in integer
cohomology. This requires analytic continuation to all singular
points.
The result is
\eqn\periods{
\Pi =\pmatrix{
-3 \partial_t \F_0\cr
t\cr
1
} =
%=\pmatrix{
%1\cr
%t\cr
%{1\over 2} t^2-{t\over 2} -{1\over 4} - \Sigma(q)}=
\pmatrix{
{1\over 2} \omega_2-{1\over 2} \omega_1-{1\over 4}\cr
\omega_{1}\cr
1}.
}
The factor of $-3$ in the above equation comes from \inter\ as we explained earlier. From above, we can read off the mirror map, giving the A-period in terms of the coordinates on the moduli space, and its inverse:
\eqn\mm{
z(Q)=Q + 6\,Q^2 + 9\,Q^3 + 56\,Q^4 +{\cal O}(Q^5)\ .
}
where $Q=e^{2 \pi i t}$, and $z$ is defined in \zphi .\foot{For the genus zero partition function this gives
$$
\partial_t \F_0=-{t^2\over 6} +{t\over 6} +{1\over 12} + 3\,Q -
\frac{45\,Q^2}{4} + \frac{244\,Q^3}{3} - \frac{12333\,Q^4}{16}+{\cal O}(Q^5),
$$
which agrees with the Gromow-Witten large radius expansion. Using this, and the definition of $\tau$ we can explicitly check \psitau . }

{}From this, we can also read off the monodromy of the periods $\Pi$
around large radius, i.e. around $z=0$ (or $\psi = \infty$). From \mm\ it follows that this is equivalent to shifting $t$ by one,
and, since
$-3 \del {\F}_0 = {1\over 2} t^2-{t\over 2} -{1\over 4} + O(e^{\pi i t})$,
this gives
\eqn\largm{
M_{\infty}=\pmatrix{
1& 1& 0 \cr
0& 1& 1 \cr
0& 0 & 1}.
}

Expanding the periods at the conifold point $\psi^3=1$, one finds the
monodromy
\eqn\conim{
M_{1}=
\pmatrix{
1& 0& 0 \cr
-3& 1& 0\cr
0& 0 & 1\cr}.}
This is the Picard-Lefshetz monodromy around the shrinking B-cycle
with intersection form \inter .  The C-period corresponds to an
auxiliary parameter, and correspondingly the C-cycle does not
intersect the $A$ and $B$ cycles.

{}From $M_{\infty}$ and $M_1$, we can recover the monodromy around
the orbifold point $M_0$, as holomorphy requires
$$M_0 M_1 M_{\infty} = 1,$$
\eqn\orbim{
M_{0}=
\pmatrix{
-2& -1& 1 \cr
3& 1& -1\cr
0& 0&1\cr } .
}
This satisfies
$(M_0)^3=1$, as it should, since the monodromy is of third order.  Note
that in all three cases, the monodromies act trivially on the C-period, which is
consistent with the fact that this corresponds simply to a parameter.
Moreover, the monodromy action on the $A$ and the $B$ periods generates the
$\Gamma_0(3)$ subgroup of $SL(2,{\IZ})$.

If instead of the $z$-plane, we choose to work with the $\psi$-plane,
then $\psi=0$ is a regular point, with trivial monodromy around it,
but instead we have three conifold singularities, at $\psi=1, \alpha, \alpha^2$, with $\alpha= e^{2 \pi i \over 3}$.
The monodromies ${\tilde M}$ in the $\psi$-plane
can be derived from the expressions above. For example,
%$$
%M^{-1}_{\psi=\infty}=
%\pmatrix{
%1& 0& 0 \cr
%3& 1& 0\cr
%3& 3& 1\cr
%}
%$$
%
$$
{\tilde M}_{1}=M_{1}, \qquad
{\tilde M}_{\alpha}=M_{0} M_{1} M^{-1}_{0}, \qquad
{\tilde M}_{\alpha^2}=M_{0}^2 M_{1} M^{-2}_{0}
$$
with monodromy at infinity given by
${\tilde M}_{\infty} = {\tilde M}_1 {\tilde M}_{\alpha} {\tilde M}_{\alpha^2}$.
These turn out to generate the ${\Gamma(3)}$ subgroup of $SL(2,{\IZ})$.
Below, we will choose to work with modular forms of ${\Gamma}(3)$,
in terms of which both $\psi$ and $z$ will be given by exactly modular forms.

\subsec{Topological Strings on Local ${\IP}^2$ and Modular Forms}

To get modular expressions for the topological string amplitudes
we need to know a bit about modular forms of the subgroup
$\Gamma(3)$ of $SL(2,\IZ)$. Essential facts about them are reviewed in Appendix A; for a detailed study of modular forms of $\Gamma(3)$, see \FarkasKra.

The set of $\theta$-constants that
generate modular forms of ${\Gamma} (3)$ is:
$$
a:=\theta^3\left[{1\over 6}\atop {1\over 6}\right], \quad
b:=\theta^3\left[{1\over 6}\atop {1 \over 2} \right], \quad
c:=\theta^3\left[{1\over 6}\atop {5\over 6}\right], \quad
d:=\theta^3\left[{1 \over 2} \atop {1\over 6}\right],
$$
which all have weight $3/2$.
They satisfy the relations \FarkasKra\
$$
c=b-a, \qquad d= a+\alpha b \ ,
$$
and the Dedekind $\eta$-function is given by $\eta^{12} = { i \over 3^{3/2}} a b c d$.
To begin with, note that since $\psi$
is a coordinate on the moduli space,
it has to be a weight zero modular form of $\Gamma(3)$.
Indeed, we find that
\eqn\psimod{
\psi(\tau)={a - c - d\over d}\ .
}
%

%Now, we can

{}From \BCOVI\ we know that the genus one free energy is given by
$$
\F_1=-{1\over 2}\log\left({\partial t\over \partial \psi}\right)-
{1\over 12}\log(1-\psi^3)\ .
$$
It is easy to show, using the $Q$-expansion of $z$ around $z=0$, that
\eqn\ptppsimod{
{\partial t \over \partial \psi}
=-\sqrt{3}{d \over \eta},
}
and that, on the other hand,
$$
\Delta = 1-\psi^3 = -3^3 {\eta^{12} \over d^4}.
$$
Combining these three expressions, we get
$$
\F_1(\tau)=-\log(\eta(\tau))+{1\over 24}\log(\Delta)\ = - {1 \over 6} \log (d \eta^3),
$$
up to an irrelevant constant term. This transforms under $\Gamma$ as $-\log(\eta)$ does, since the discriminant $\Delta$ is invariant, which is exactly what we predicted.
As a consistency check, if we use the $Q$-expansion of $q$ and the modular expression for $\F_1$ we get the expansion
$$
\F_1 = - {1 \over 12}\log Q + {Q \over 4} - {3 Q^2 \over 8}-{23 Q^3 \over 3} + {\cal O}(Q^4),
$$
which is precisely the genus $1$ amplitude of local $\IP^2$.

\subsec{Higher Genus Amplitudes}
To find the higher genus amplitudes, we need the modular expression for the
Yukawa coupling $C_{ttt}={\partial^3  \over \partial t^3}\F_0$.
We know that
$$
C_{ttt} = -{1 \over 3}{\partial \tau \over \partial t} = -{1 \over 3}{\partial \psi \over \partial t}{\partial \tau \over \partial \psi}.
$$
Using the modular expressions for $\psi$ \psimod, for ${\partial t \over \partial \psi}$ \ptppsimod, and the formulae for logarithmic derivatives derived in Appendix A, we get
\eqn\Yukptwo{
C_{ttt} = -{1 \over 3^{5/2}} {d \over \eta^9}.
}
Another useful object is the $\Gamma(3)$-invariant Yukawa coupling, expressed in terms of the globally defined variable $\psi$. We obtain
\eqn\Cpsi{
C_{\psi \psi \psi} = \left({\del t \over \del \psi} \right)^3 C_{ttt} = - {9 \over \Delta}.
}

Using the results of the previous subsection, we can now find a modular
expression for higher genus amplitudes, through their Feynman expansions. The propagator
$E(\tau)$
must transform under modular transformations as in \Etrans\
$$
E((A\tau + B)/(C\tau+D)) = (C\tau+D)^2 E(\tau) - 3 \,C\, (C\tau+D);
$$
the factor of $-3$ comes from the intersection numbers \inter .
For example, we can take
$$
E = -{2 \pi i \over 4} E_2 (\t).
$$
We could have worked with the full $E'=6
{\partial \over \del \tau} \F_1$ as well, since the propagator is defined up to a modular invariant piece; it would have only changed
the modular invariant $h_2^{(0)}$.

We obtain that the general form of the higher genus amplitudes reads
\eqn\genformptwo{
\F_g= X^{g-1}\sum_{k=0}^{3 (g-1)} {E}_2^{k} h_g^{(3 g-3-k)}(K_2,K_4,K_6)
}
where we defined the weight $-6$ object
$$
X = {d^2 \over 2^9 3^4 \eta^{18} }={1 \over 1536} C_{ttt}^2
$$
and the ring of modular forms of $\Gamma(3)$ generating the weight $2d$ forms $h_g^{(d)}$
is given by
$$
K_2= -\alpha^2 {(a-\alpha c)^2\over \eta^2},
\qquad K_4={1\over \alpha^2-1} {ac (a+c)(\alpha^2a -c)\over \eta^4},
\qquad K_6={(ac)^2(a+c)^2\over \eta^6}.
$$
The coefficients of $E_2$
are fixed by the Feynman graph expansion and we obtain for example
\eqn\higenpII{
\eqalign{
h_2^{(0)} &= \F_2 - X \biggl( 5 E_2^3 +  E_2^2 K_2 + {1 \over 3} E_2 K_2^2 \biggr),
\cr
h_3^{(0)} &= \F_3 - X^2 (
180  E_2^6 + 240  E_2^5 K_2 + 4  E_2^4 (145K_2^2 - 1008 K_4)
\cr & +
{32 \over 9}  E_2^3 (199 K_2^3 - 1908K_2 K_4 + 648 K_6)+
{4\over 5} E_2^2 (563 K_2^4 - 7936 K_2^2 K_4 + 26496 K_4^2)
\cr &
+{16 \over 15} E_2 (149 K_2^5 - 2536 K_2^3 K_4 + 11952 K_2 K_4^2 - 3456 K_4 K_6)).}}

Now, using known results for $\F_g$ in the large radius limit
(obtained for instance through the topological vertex formalism), we
can find the $h_g^{(0)}$'s explicitly --- this corresponds to fixing the holomorphic
ambiguity in the BCOV formalism. For instance, we obtain

\eqn\pambig{
\eqalign{
h_2^{(0)} &= {11 \over 69120} + {1 \over 34560 \Delta} - {1 \over 7680 \Delta^2},
\cr
h_3^{(0)} &= {17\over  6289280}+{269\over 46448640\Delta}-
{19393\over 278691840 \Delta^2}+
{337\over 2211840 \Delta^3} -
{373\over 4128768 \Delta^4} .}
}

\subsec{The $\,{\IC}^3/{\IZ}_3$ Orbifold Point}

In this section we explain how to extract the Gromov-Witten generating functions of
the orbifold $\,{\IC}^3/{\IZ}_3$ from the large radius amplitudes, through the wave function formalism.

Let us first discuss this theory from the mirror A-model point of
view.  The target space $X$ is an $X={\IC}^3/{\IZ}_3$ orbifold, with
${\IZ}_3$ acting on the three coordinates $z_{i}$, $i=1,2,3$ by
$$
z_i \rightarrow \alpha z_i, \qquad \alpha = e^{2\pi i \over 3}.
$$
In quantizing string theory on $X$, the Hilbert space splits into $3$
twisted sectors, corresponding to strings closed up to
$\alpha^{k}$, $k=0,1,2$ (and projecting onto ${\IZ}_3$ invariant
states). The supersymmetric ground states in the $k$-th sector
correspond to the cohomology of the fixed point set of $\alpha^k$.
This has an interpretation in terms of the cohomology of $X$ as well. In
the case at hand, the ground states in the sector twisted by
$\alpha^k$ correspond to the generators of $H^{k,k}(X)$. Namely,
the contribution to the cohomology of $X$ is determined by the $U(1)_L
\times U(1)_R$ charges of the states, where the charge $(p_i,q_i)$
corresponds to $H^{p_i,q_i}$. In the twisted sectors, however, these
receive a zero-point shift: in the sector twisted by $z_i \rightarrow
e^{2\pi i k_i} z_i$ with $0\leq k_i <1$ the shift is $(\sum_i k_i,
\sum_i k_i)$.
As there
is precisely one such state for each $k$,
the stringy cohomology of the orbifold agrees with
the cohomology of the smooth resolution of $X$, i.e. the
$O(-3) \rightarrow {\IP^2}$, as is generally true (see however \refs{\orbifwithdiscretetorsion}).

As explained in \refs{\quantumsymmetries}, the
orbifold theories have discrete $quantum$ symmetries.
In the present case, this is the $\IZ_3$ symmetry
which sends a state in the $k$'th twisted sector to itself
times $\alpha^k$.  This is respected by interactions,
so it is a well defined symmetry of the quantum theory.
This implies that the only
non-vanishing correlation functions are those
that have net charge zero (mod $3$). In particular, if we
consider correlation functions of $n$ insertions of
topological observables ${\cal O}_\sigma$
corresponding to the generator of $H^{1,1}(X)$,
$$
\langle\; \underbrace{{\cal O}_\sigma \,{\cal O}_\sigma
\ldots {\cal O}_\sigma}_n \; \rangle_g
$$
at any genus $g$, this does not vanish only if $n=0 \; ({\rm mod} \;3)$
We will describe in this section how to compute the
generating functions of correlation functions at genus $g$
$$
\rF_g(\sigma) = \sum_{n} {1\over n!}
\, \langle \,({\cal O}_\sigma)^n\,\rangle \, \sigma^n
$$
and show that this is indeed the case. By $\rF_g$ here,
we mean the generating function at the orbifold point ---
in this section, we will denote the generating function in the large radius limit by $\iF_g$ to avoid confusion.

{}From what we explained in section 3, the expectation is the
following. The good coordinate in one region of the moduli space
generally fails to be good at other regions of the
moduli space.
The good variable at large radius is
$t$, as the corresponding monodromy is trivial \largm , according to our criterion in section 3.
However, the monodromy of the period $t$ is not trivial around the orbifold point,
being given by \orbim ,
as $3\neq 0$.  Correspondingly, even though we know the
topological string amplitudes near the large radius point, we cannot
simply analytically continue them to the orbifold point --- the resulting objects would have bad singularities.
Changing to good variables at the orbifold point involves a
wave function transform that mixes up the genera.

What is the good variable at the orbifold point?
Clearly, it is the mirror B-model
realization of the parameter $\sigma$
that enters the orbifold Gromov-Witten partition functions in the
A-model language and corresponds to $H^{1,1}(X).$ The dual
variable $\sigma_D$
$$
\sigma_D = -3 {\del\over \del \sigma} \rF_0
$$
corresponds to $H^{2,2}(X)$. To identify them in the B-model,
note that, on the one hand,
under the quantum symmetry ${\IZ}_3$ symmetry
$\sigma$ and $\sigma_D$ transform as
$$
(1, \sigma,  \sigma_D) \rightarrow ( 1, \alpha \, \sigma,  \alpha^2 \,
\sigma_D).
$$
On the other hand,
the symmetry acts in the mirror theory by \quantumsymmetries\
$$
\psi \rightarrow \alpha \, \psi.
$$
The fixed point of this, $\psi =0$, corresponds to the elliptic curve
with $\IZ_3$ symmetry, which is mirror to the $\;\IC^3/\IZ_3$ orbifold.
We can easily find the solutions to the Picard-Fuchs equations with these properties.

A basis of solutions is given by the
hypergeometric system $_3F_2$
\eqn\Bk{
B_k(\psi)= {(-1)^{{k\over3}} \over k} (3 \psi)^k\sum_{n=0}^\infty
{\left(\left[ {k \over 3}\right]_n\right)^3\over
\prod_{i=1}^3 \left[{k+i\over 3}\right]_n }\psi^{3n}\ ,
} 
for $k=1,2$,
where we defined the Pochhammer symbols $[a]_n:={\Gamma(a+n)\over \Gamma(a)}$.
We also set $B_0 (\psi)= 1$.
The $B$'s diagonalize the monodromy around
the orbifold point, namely $\psi\rightarrow \alpha \psi$ takes
$$
(B_0,B_1,B_2) \rightarrow (B_0, \alpha B_1,\alpha^2 B_2).
$$
Consequently, we can identify
$$
(1, \sigma,\sigma_D)= (B_0,B_1,B_2).
$$
The relative normalization of $\sigma$ and $\sigma_D$ can be fixed
using $\sigma_D = - 3 {\del \rF_0 \over \del \sigma}$ and hence ${\del
{\tilde \tau} \over \del \sigma} = {\del^2 \sigma_D \over \del
\sigma^2} = - 3 C_{\psi \psi \psi} \left({\del \psi \over \del \sigma}
\right)^3$, since $\psi$ is globally defined.

We can already make a prediction for the genus zero free energy at the orbifold point, up to an overall 
constant. By integrating $\sigma_D = \sign 3 {\partial \rF_0 \over \partial \sigma}$, we get
$$
\rF_0(\sigma) = \sum_{n=1}^\infty {\rN_{g=0,n} \over (3n )!} \sigma^{3n}
$$
where, for example
$$
\eqalign{
\rN_{0,1} =& {1\over 3}, \;\; \rN_{0,2} = -{1\over 3^3}, \;\; \rN_{0,3} = {1\over 3^2},
 \;\; \rN_{0,4}=-{1093 \over 3^6},\cr
&\;\; \rN_{0,5} = {119401 \over 3^7}, \;\; \rN_{0,6} = -{27428707 \over 3^8},\ldots}
$$

Let us now turn to higher genus amplitudes. The analytic continuation from the point at infinity
to the orbifold point can be done with the Barnes integral, as also
explained in
\Candelas. This relates
\eqn\basisch{
\Pi =\pmatrix{
-{1\over 1-\alpha} c_2
&{\alpha\over 1-\alpha} c_1 & {1\over 3}
\cr
c_2& c_1 &0 \cr
0&0&1} \pmatrix{\sigma_D \cr \sigma \cr 1
}}
with the
coefficients
\eqn\defc{
c_1=\sign {i\over 2 \pi} {\Gamma\left({1\over 3}\right)\over
\Gamma^2\left({2\over 3}\right)}, \qquad c_2=\sign -{i \over 2 \pi}
{\Gamma\left({2\over 3}\right)\over \Gamma^2\left({1\over 3}\right)\ },
}
which are not integers. This is because
the natural basis $(\sigma, \sigma_D)$ diagonalizes the monodromy around the orbifold point, and this cannot be
done in $SL(2,{\IZ})$.\foot{We could have derived the change of basis
in another way.  There is another natural basis at the orbifold,
$(C_0,C_1,C_2)$, corresponding to the 3 fractional branes.  This basis has monodromy around the orbifold point,
which is the cyclic ${\IZ_3}$ permutations of the branes,
$$
\pmatrix{C_0 \cr C_1\cr C_2} \rightarrow
\pmatrix{0&1&0\cr
0&0&1\cr
1&0&0}\pmatrix{C_0 \cr C_1\cr C_2}.
$$
The fractional brane basis is related to the large radius basis by an integral
transformation --- respecting the integrality of the D-brane charges ---
and the symplectic form.
On the other hand, it is known \refs{\DouglasMoore} how the fractional
branes couple to the twisted sectors: in particular, the $i$-th twisted
sector corresponds to $\sum_j \alpha^{ij} C_j$. This reproduces
\basisch .}
Note that $c_1 c_2 = {\alpha(\alpha - 1) \over (2 \pi i)^3}$; correspondingly
the change of basis $does$ $not$ preserve the symplectic form,
we have rather that
$$
dp \wedge dx = {1\over \beta} d\sigma_D\wedge d\sigma
$$
where
$$
\beta = -{(2\pi i)^3}.
$$
Because of this fact, the analysis of
section 2 goes through, but one has to be careful with
normalizations. More precisely, it implies that the effective string coupling
at the orbifold $(g_s^{\rm orb})^2$ is renormalized relative to
the large radius $g_s^2$ by $(g_s^{\rm orb})^2 = \beta g_s^2$.

Then, knowing the Gromov-Witten amplitudes
at large radius, we can predict them at the orbifold:
\eqn\oneorbi{
\beta^{g-1}\rF_g =  \iF_g  + \Gamma_g(\Delta,~\del_{i_1}\ldots \del_{i_n} {\cal F}^{\infty}_{r<g}),}
where the coefficient $\beta$ comes from the renormalization of the string coupling, and
$$
\Delta = {3\over \tau + C^{-1} D}.
$$
The coefficient $3$ above comes from \inter . The coefficients
$C$ and $D$ are computed from (the inverse of) \basisch\ as before,
which gives
\eqn\compl{
{C^{-1} D} = {1\over 1-\alpha}.
}
In order to extract the $\sigma$-expansion of $\rF_g$ such as we presented for $\rF_0$,
we compute the right hand side of \oneorbi\ in terms of the period $t$, and then use the relation
between $\sigma$ and $t$ given in \basisch\ to get expansions around $\sigma=0$.

{}Since ${\tilde \tau} = {\del \sigma_D \over \del \sigma}$ vanishes
at the orbifold point $\sigma=0$,
it follows from \basisch\  that
\eqn\tausig{
\tau(\sigma=0) = {\alpha \over 1 - \alpha}.
}
Numerically, this corresponds to $q (\sigma = 0 ) = -
e^{-{\pi \over \sqrt{3}}} \sim -0.16$; at this value, the
$q$-expansion of the modular expression \genformptwo\ still converges
rapidly. Indeed, since the coefficients of the $\sigma$-expansion of
the topological string amplitude at the orbifold point are rational
numbers, they can be easily recovered from their convergent
$q$-expansion.

At genus $1$, we get
$$
\rF_1(\sigma) = \sum_{n=1}^\infty { \rN_{g=1,n} \over (3n )!} \sigma^{3n}
$$
where, for instance,
$$
\eqalign{
\rN_{1,1} = 0,~~\rN_{1,2}={1\over 3^5},~~\rN_{1,3}=-{14\over 3^5},\cr
\rN_{1,4}={13007\over 3^8},~~\rN_{1,5}=-{8354164\over 3^{10}},\ldots
}
$$
It is good to note that simply expanding
$\iF_g (\tau)$ near $\tau(\sigma =0) $, that is, doing only the analytic continuation of the amplitudes,
would lead to
non-rational coefficients in the $\sigma$-expansion.

Instead of \oneorbi\ , it is
faster to use the recursion relations at the orbifold
point directly in terms of the modular ambiguity \pambig\
and the corresponding propagator,
$$
E^{orb}(\tau) =
\lim_{{\bar \tau} \rightarrow {\bar \tau}(\sigma=0)}
{\hat E}(\tau,\bar{\tau})
$$
where
$$
{\bar \tau}(\sigma=0) =
-C^{-1}D
$$
is just the complex conjugate of \tausig .
This follows from the fact that ${\hat \F}_g$, on the one hand,
satisfies the same recursion relations as $\iF_g$ with $E$'s
and $\iF_r$'s replaced by their hatted counterparts,
and on the other hand ${\hat \F}_g(\tau, {\bar \tau})$ at ${\bar \tau}$ set
to ${\bar \tau}=-C^{-1}D$ gives $\rF_g$. In fact, the right
hand side of \oneorbi\ can be interpreted as computing just that.
Either way, for $\rF_g$, we find that
$$
\rF_g(\sigma) = \sum_{n=0}^\infty {\rN_{g,n} \over (3n )!} \sigma^{3n}
$$
with the numbers $\rN_{g,n \geq 1}$
\vskip 5 mm
{\vbox
{\ninepoint{
$$
\vbox{\offinterlineskip\tabskip=0pt
\halign{\strut
\vrule#&
&\hfil ~$#$
&\hfil ~$#$
&\hfil ~$#$
&\hfil ~$#$
&\hfil ~$#$
&\hfil ~$#$
&\vrule#\cr
\noalign{\hrule}
&g
&n=1
&2
&3
&4
&5&\cr
\noalign{\hrule}
&0
&{1 \over 3}
&-{1 \over 3^3}
&{1 \over 3^2}
&-{1093 \over 3^6}
&{119401 \over 3^7}&\cr
&1
&0
&{1\over 3^5}
&-{14\over 3^5}
&{13007\over 3^8}
&-{8354164\over 3^{10}}&\cr
&2
&{1\over 2^4\cdot 3^4 \cdot 5}
&-{13\over 2^4\cdot3^6}
&{20693\over 2^4\cdot 3^8 \cdot 5}
&-{12803923\over 2^4 \cdot 3^{10}\cdot 5}
&{31429111\over 2^4 3^{10}}&\cr
&3
&-{31\over 2^5 3^5 5 \cdot 7}
&{11569\over 2^5 3^9 5\cdot 7}
&-{2429003 \over 2^5 3^{10} 5\cdot 7}
&{871749323\over 2^4 3^{11}5\cdot 7}
&-{1520045984887\over 2^5 3^{13} 5\cdot 7} &\cr
&4
&{313\over 2^7 3^9 5^2}
&-{1889\over 2^7 3^{9} }
&{115647179\over 2^6 3^{13} 5^2 }
&-{29321809247\over 2^8 3^{12} 5^{2}}
&{22766570703031\over  2^7 3^{15} 5}&\cr
&5
&-{519961\over 2^9 3^{11} 5^2 7 \cdot 11}
&{196898123\over 2^9 3^{12} 5^2 7\cdot 11}
&-{339157983781\over 2^9 3^{14} 5^2 7\cdot 11}
&{78658947782147 \over 2^9 3^{16} 5 \cdot 7}
&-{1057430723091383537\over 2^9 3^{17} 5^2 7 \cdot 11}&\cr&
6
&{ 14609730607\over 2^{12} 3^{13} 5^3 7^2  11}
&-{258703053013\over 2^{10} 3^{15} 5^1 7^2 11}
&{2453678654644313\over 2^{12} 3^{14} 5^3 7^2 11}
&-{40015774193969601803 \over 2^{11} 3^{18} 5^3 7^2 11}
&{5342470197951654213739\over 2^{12} 3^{19} 5\cdot 7^2 11}&\cr
\noalign{\hrule}}\hrule}$$}}
where we also included the genus $0$ and $1$ numbers obtained earlier for completeness.

The $n=0$ numbers, corresponding to untwisted maps for $g \geq 2$ (these are not well-defined for $g=0,1$), read
$$
\eqalign{
\rN_{2,0} =& {-1 \over 2160} + { \chi \over 5760}, \;\; \rN_{3,0} = {1\over 544320} - {\chi \over 1451520}, \;\; \rN_{4,0} = -{7\over 41990400}+{\chi \over 87091200},\cr
\rN_{5,0} =& {3161 \over 77598259200} - {\chi \over 2554675200}, \;\; \rN_{6,0} = -{6261257 \over 317764871424000} + {691 \chi \over 31384184832000},\ldots}
$$
where $\chi$ is the ``Euler number" of local $\IP^2$. The natural value of $\chi$ is $3$.

Generally in Gromov-Witten theory the denominators come from dividing
by the finite automorphisms of the moduli space ${\cal M}_{g,n}$. In
the $\IZ_3$ orbifold case there are obviously various automorphisms of
order $3$, corresponding to the powers of $3$ in the denominators. We
note that all other prime factors in the denominators do not exceed
the prime factors in $|B_{2g} B_{2g-2}| \over 2g (2g-2) (2g-2)!$. Automorphism groups of
this order arise already for the constant map Gromov-Witten invariant.

%%%%%%%%%%%%
%%%%%%%%%%%%
\newsec{Local $\IP^1 \times \IP^1$}

Our last example is
the Gromov-Witten theory of the Calabi-Yau $Y$ which is the
total space of the canonical bundle over $\IP^1 \times \IP^1$.
We will study this using modularity of the $B$-model topological string
on the mirror manifold $X$.

To start with, let us review elementary facts about $Y$.
Let $A_1$ and $A_2$ denote the classes
of the two $\IP^1$'s in $H_2(Y)$. There is also one compact four cycle --
the $\IP^1 \times \IP^1$ itself, and denote by $B$ the corresponding class
in $H_4(Y)$.
The intersection numbers of the cycles on $Y$ are
$$
\#(A_1  \cap B) = -2  = \# (A_2 \cap B).
$$
The class $C = A_1 - A_2$ does not have a dual cycle in $H_4(Y)$, as it does not intersect $B$. From our discussion in section 2, $C$ will correspond to a {\it non-normalizable} modulus of the theory. For the normalizable modulus $A$
we can take $A_2$, for example, so let us define
$$
A = A_2, \qquad C = A_1 - A_2.
$$

The mirror manifold is a family of elliptic curves $\Sigma$, which
is given by the following equation
\refs{\ChiangTZ, \HoriVafa} in $\IP^1 \times \IP^1$:
\eqn\mirrorpp{
x_0^2 y_0^2 + {z_1} x_1^2 y_0^2 + x_0^2 y_1^2 + {z_2} x_1^2 y_1^2 +
x_0 x_1 y_0 y_1 = 0,
}
where $[x_0:x_1]$ and $[y_0:y_1]$ are homogeneous coordinates of the two
$\IP^1$'s. The large radius point corresponds to $z_1=0=z_2$.

Let $t_1$ and $t_2$ denote the
periods of the one form $\lambda$ around the 1-cycles mirror dual to $A_1$ and
$A_2$ (which we also denote by $A_1$ and $A_2$):
$$
t_1 = \int_{A_1} \lambda,  \qquad t_2 = \int_{A_2} \lambda,
$$
and let $t_D$ be the period around the 1-cycle mirror dual to $B$:
$$
t_D = \int_B \lambda.
$$
The periods $t_1$ and $t_2$ compute the physical K\"ahler parameters,
i.e. the masses of BPS D2-branes wrapping the two
${\IP}_1$'s.\foot{The ${\IP}^1$'s of the embedding space of the mirror
will hopefully not be confused with the two ${\IP}^1$'s generating $H_2(Y)$ on the $A$-model side.}
At large radius the complex structure
parameters $z_1$ and $z_2$ are related to the
K\"ahler parameters $t_1$, $t_2$ of $Y$ by
$$
z_{1,2} \sim e^{2\pi i t_{1,2}}.
$$
More specifically, we can find the periods $t_i$ in terms of the parameters $z_i$ as the solutions of the Picard-Fuchs equations of $X$
\eqn\PFpone{
\eqalign{
{\cal L}_1 &= \Theta_1^2 - 2z_1 (\Theta_1+ \Theta_2)(1 + 2\Theta_1 +2 \Theta_2),\cr
{\cal L}_2 &= \Theta_2^2 - 2z_2 (\Theta_1+ \Theta_2)(1 + 2\Theta_1 +2 \Theta_2),
}
}
where $\Theta_i = z_i {\partial \over \partial z_i}$ for $i=1,2$.
The solutions around the large radius point $z_1=0=z_2$ can be determined by the Frobenius method from
$$
\omega(z_1,z_2,r_1,r_2):=\sum_{m,n=1}^\infty {z_1^{r_1+m} z_2^{r_2+n} \over \Gamma(-2(m+r_1)-2(n+r_2)+1)\Gamma^2(m+r_1+1)\Gamma^2(n+r_2+1)}
$$
as
$$
t_{i}={1\over (2 \pi i)}\left. {\dd\over \dd r_i}
\omega(z_1,z_2,r_1,r_2)\right|_{r_{1,2}=0}.
$$
Thus
$$
t_{1}(z_1,z_2) = \log (z_1) + 2z_1 + 2z_2 + 3 z_1^2 + 12 z_1 z_2 +
3 z_2^2 + \ldots
$$
and similarly for $t_2$ with $z_1$ and $z_2$ exchanged.
By inverting the above, we get the mirror maps:
\eqn\mirrorpone{
\eqalign{
z_1 &= q_1 -2( q_1 + q_1 q_2) +3 (q_1^3 + q_1 q_2^2) -4 ( q_1^4 +q_1^3 q_2+q_1^2 q_2^2 + q_1 q_2^3)+\ldots\cr
z_2 &= q_2 -2( q_2 + q_1 q_2) +3 (q_2^3 + q_2 q_1^2) -4 ( q_2^4 +q_2^3 q_1+q_1^2 q_2^2 + q_2 q_1^3)+\ldots
}
}
where $q_i = \exp (2 \pi i t_i)$ for $i=1,2$.

In addition to this there are two other solutions to the Picard-Fuchs equations. First, there is a double logarithmic solution, which is the period $t_D$ introduced previously. Second, there is a constant solution, corresponding to the period mirror to the
D0 brane in the A-model. This constant period, together with
$$
m=t_1-t_2 = \int_{C} \lambda,
$$
where $C$ is the 1-cycle of the curve mirror dual to the class $C$ of $Y$ (again we use the same letter to denote mirror dual objects),
should be regarded as constant parameters that enter in specifying the
model. In fact, it is easy to
see that the period $m$ does not
receive instanton corrections, i.e.
$q_m =\exp (2 \pi i m)$ satisfies
$$
q_m = q_1/q_2 = z_1/z_2,
$$
which is consistent with the
interpretation of $m$ as an auxiliary parameter.

In the following we will denote the physical modulus by $T$
$$
T = t_2 = \int_{A} \lambda,
$$
and define $Q = \exp (2\pi i T)$.

In order to find the modularity properties of the amplitudes, we now
study in more detail the family of elliptic curves $\Sigma$.

\subsec{The Family of Elliptic Curves}

The family of elliptic curves $\Sigma$ in \mirrorpp\ can be brought into Weierstrass form,\foot{To do so, we first use the Segre embedding of
$\IP^1 \times \IP^1$ into $\IP^3$ given by the map
$$
([x_0: x_1],[y_0:y_1]) \mapsto [X_0:X_1:X_2:X_3]=[x_0 y_0, x_1 y_0, x_0 y_1, x_1 y_1],
$$
where $[x_0:x_1]$ and $[y_0:y_1]$ are homogeneous coordinates of the two $\IP^1$'s and $X_i$, $i=0,\ldots,3$ are homogeneous coordinates of $\IP^3$. Then $\IP^1 \times \IP^1$ is given by the hypersurface
\eqn\embpone{
X_0 X_3 - X_1 X_2 = 0
}
in $\IP^3$. The family of elliptic curves $\Sigma$ is now given by the complete intersection of \embpone\ and the hypersurface defined by
\eqn\embcur{
X_0^2 +z_1 X_1^2 + X_2^2 + {z_2 } X_3^2 + X_0 X_3 = 0 .  } After a
linear change of variable, \embcur\ becomes linear in $X_3$, so $X_3$
can be eliminated from \embcur\ and \embpone\ to get a cubic equation
in $\IP^2$. Then, given any cubic in $\IP^2$ we can use Nagell's algorithm
\refs{\Cassels, \Connell} to transform it into its Weierstrass form.}
$$
y^2 = 4x^3 - g_2 x - g_3
$$
with
$$
\eqalign{
g_2 &= {2^{2/3} \over 3} (16 z_1^2 + (1-4z_2)^2 + 8z_1 (-1+28z_2) ),\cr
g_3 &= {2 \over 27} (64 z_1^3+(-1 + 4z_2)^3 - 48 z_1^2 (1+44 z_2) + z_1 (12 + 480 z_2 - 2112 z_2^2) .
}
$$
Its $j$-function reads
\eqn\jpone{
j (\tau) = { (16 z_1^2 + (1-4z_2)^2 + 8 z_1 (-1 + 28 z_2))^3 \over z_1 z_2 (16 z_1^2 + (1-4 z_2)^2 - 8 z_1(1+4 z_2))^2}.
}
As usual, by $j (\tau)$ we mean that the $j$-function is a function of the
standard complex parameter $\tau$ of the family of elliptic curves
$\Sigma={\IC / (\IZ \oplus \tau\IZ )}$.

%or, letting $\Delta$ be the discriminant of $C$,
%$$
%j(\tau) = { (\Delta +192 z_1 z_2)^3 \over 1728 z_1 z_2 \Delta^2}.
%$$
%
As it turns out, we have met this curve before! Recall that the $j$-function of the $\Gamma(2)$ modular curve, the $SU(2)$ Seiberg-Witten curve, is \jsw\
\eqn\jsw{
j(\tau) = { 64 (3+u^2)^3 \over (u^2-1)^2}.
}
If we make the substitution
\eqn\zsub{
u = {q_m^{-1/2} \over 8 z_2} - {1 \over 2}( q_m^{1/2}+q_m^{-1/2} )
}
in \jsw, we get exactly the $j$-function \jpone, using the fact that
$q_m = z_1/z_2$. Since the $j$-function captures all the
coordinate-invariant data of the elliptic curve, the curves in the family mirror to local $\IP^1 \times \IP^1$ are in
fact isomorphic to the curves in the $SU(2)$ Seiberg-Witten family, through reparameterization of the moduli space
as in \zsub. In particular, it follows immediately that
the curves in the family $\Sigma$ have monodromy group
$\Gamma(2)$.

We could also have found the monodromy transformations of the
periods directly from the Picard-Fuchs equations, as we did for local $\IP^2$, but it
requires more work. The $j$-function approach,
when the mirror geometry is a family of elliptic curves,
provides a simpler way to determine the monodromy group,
at least the part of it restricted to the physical periods.
Fortunately, this is all that is relevant for our purposes.

Using this result, we can borrow heavily the results from the $SU(2)$ theory.
In particular, using the expression for $u$ in terms of modular
forms of $\Gamma(2)$ in
\utau\ and relating $z_2$ to the period $T$, we find\foot{Note that we could invert the series because $q_m$ is just a parameter, i.e. it must be $\tau$-independent.}
\eqn\Qtau{
Q(q_m,q)= q_m^{-1/2} q^{1/2} - (2+2q_m^{-1})\; q + q_m^{-3/2}(5-4q_m+5q_m^2)\; q^{3/2} +\ldots
}
where $q = e^{2\pi i \tau}$, $q_m = e^{2 \pi m}$ and $Q = e^{2 \pi i
T}$. {}From this expansion, we see that the period $T$ does not only
depend on $\tau$; the coefficients of the power series in $q$ depend
explicitly on the auxiliary parameter $m$ (or $q_m$).

\subsec{Genus $0$, $1$ and Yukawa Coupling}

Let us start by finding the partition function at genus $1$. Recall
that $\F_1$ is fixed by its modular properties and its behavior at the
discriminant of the family of elliptic curves $\Sigma$.
In the local $\IP^1 \times \IP^1$ case, we can show that
\eqn\fonepone{
\F_1 = - \log \eta(\tau)
}
transforms as required and has precisely the good behavior at the discriminant --- this is the same expression as in $SU(2)$ Seiberg-Witten theory. As a consistency check, if we expand \fonepone\ using the expansion of $q$ in terms of $q_m$ and $Q$ we get
$$
\eqalign{
\F_1 =& -{1 \over 24} \log (q_m Q^2) - {1 \over 6}(1+q_m)Q -
{1 \over 12} (1 + 4 q_m + q_m^2) Q^2 \cr
&~~~- {1 \over 18} ( 1 + 9 q_m + 9 q_m^2 +q_m^3 )Q^3 +{\cal O}(Q^4),
}
$$
which reproduces precisely the genus one
partition function of local $\IP^1 \times \IP^1$.

Now consider the Yukawa coupling, i.e.
the
third derivative of $\F_0(m,T)$ with respect to $T$,
which we will need to compute higher genus amplitudes.
Using
$$
{\partial^3
\over \partial T^3} \F_0(m,T) = -
{1 \over 2} {\partial
\over \partial T} \tau(m,T)
$$
and the expansion for $\tau$ in terms of $q_m$ and $Q$ we get the following expansion
\eqn\Fzeropone{
{\partial^3
\over \partial T^3} \F_0(m,T)
= -1 -2 (1 + q_m)Q
-2 (1 + 16 q_m + q_m^2) Q^2+ {\cal O}(Q^3).
}
However, what we would like to obtain is a modular expression for
${\partial^3 \over \partial T^3}
\F_0 $
defined globally over the moduli
space of complex structures, such as our expression \fonepone\ for
$\F_1$, not just an expansion in the large complex structure limit.

To identify the modular form we make use of the change of variable
\zsub, which relates the usual $\Gamma(2)$ curve to our curve with the
auxiliary parameter $q_m$. Through this change of variable, we identify the $j$-functions of the two curves, and correspondingly the parameter $\tau$, via the $q$-expansion of the $j$-function. In particular, this implies a relation between the periods
$$
a = a(T,m),
$$
where $a$ is the usual Seiberg-Witten period, coming from the identification of the $j$-functions, which we write schematically as
$$
j(a) = j(\tau) = j(T,m).
$$
As a result, acting on any function of $\tau$ (at fixed $m$), we get
$$
{\del \over \del T} = {\del a \over \del T} {\del \over \del a}.
$$
For instance, we can write
$$
{\partial^3 \F_0
\over \partial T^3}
= -{1 \over 2} {\partial \tau
\over \partial T}
 = -{1 \over 2} {\partial a \over \partial T}{\partial \tau
\over \partial a}.
$$
Now, we saw in section 5 that
$$
{\del \tau \over \del a} = - 2 {\sqrt{b}\over c d}(\tau)
$$
and we can compute that
\eqn\fexpr{ f := {\partial a \over \partial T} = - {1 \over 2} \left(
q_m^{1/2} + q_m^{-1/2} + 2 {d + c \over b} (\tau)\right)^{1/2} }
using \Fzeropone\ and \Qtau . In the above equations we used the
modular forms $b,c$ and $d$ as defined in the $\Gamma(2)$ part of
Appendix A. Putting all this together, we get
$$
{\partial^3 \F_0
\over \partial T^3}
=
-{1 \over 2}{\sqrt{b} \over c d} \left( q_m^{1/2} + q_m^{-1/2} + 2 {d + c \over b} (\tau)\right)^{1/2}
$$
which is a modular form of $\Gamma(2)$ of weight $(-3)$, as expected.
Note that $f$ itself has weight zero.

To summarize, given the function $f = {\del a \over \del T}$ in
\fexpr , which relates the $a$-period of
the $\Gamma(2)$ curve to the $T$ and $m$ periods of the $\IP^1 \times
\IP^1$ curve, we directly obtain modular expressions for the higher
genus amplitudes in terms of the modular expressions already obtained
for $SU(2)$ Seiberg-Witten theory.

\subsec{Higher genus amplitudes}

First, we can take the propagator to be
$$
E(\tau) = -{2 \pi i \over 6} E_2 (\tau),
$$
 which is the same propagator as in $SU(2)$
Seiberg-Witten theory, up to a sign (see section 5). The sign comes from the different conventions for the relative orientation of the A and the B-cycles.

To get higher genus amplitudes, we use the by now familiar Feynman expansions with the above propagator. To relate the expansions to the $SU(2)$ Seiberg-Witten expansions, we simply use the chain rule for derivatives: whenever we need to take derivatives with respect to $T$ in the Feynman expansions, we use the function $f$ given in \fexpr\ to write
$$
{\partial \over \partial T} = f {\partial \over \partial a}.
$$
This relates the amplitudes on
the local
${\IP^1\times \IP^1}$ to those in the $SU(2)$ Seiberg-Witten theory,
up to an exactly modular form.
Plugging all these results in the Feynman expansion for the genus 2 partition function $\F_2$ we get the nice and simple expression for the modular function $h_2^{(0)}$ in terms of the partition functions $\F_g^{\rm SW}$, $g \leq 2$ of $SU(2)$ Seiberg-Witten theory:
$$
h_2^{(0)} =\F_2 + {1 \over 4}\F_2^{\rm SW} \left(q_m^{1/2}+q_m^{-1/2}+ 2 {c + d \over b}\right) - {1 \over 576}{ E_2^2 \over c d}.
$$

This is an interesting result. Through our modular formalism, we can
express higher genus amplitudes of local Calabi-Yau manifolds in a
very simple way in terms of higher genus amplitudes of the
corresponding theory with no auxiliary parameters --- in this case
$SU(2)$ Seiberg-Witten theory. More precisely, given two theories
governed by elliptic curves with $j$-functions related by a change of
variables (that generically also involves the auxiliary parameters),
all one needs to do is to determine the function $f = { \partial a
\over \partial T}$ relating the physical periods, and everything else
follows from the formalism.

Finally, by plugging in the known expansion for $\F_2$ (obtained for instance through the topological vertex formalism) we could determine $h_2^{(0)}$, and show that it is a modular form of weight $0$, as we did for local $\IP^2$. We could also go to higher genera, and relate the expressions to the Seiberg-Witten expressions; we will not present the explicit formulae here, but it is straightforward to calculate them.

\subsec{Seiberg-Witten Limit}

Let us end this section by showing that the double scaling limit to recover $SU(2)$ Seiberg-Witten theory from the local $\IP^1 \times \IP^1$ topological string amplitude is consistent with our results above. Since we know the $j$-function of the mirror family of elliptic curves in terms of the complex moduli $z_1$ and $z_2$, we first express the limit in these parameters, and then show that taking the limit gives the $j$-function of the $SU(2)$ Seiberg-Witten curve.

The double scaling limit was explained in details in \refs{\KachruFV, \KatzFH}. Define first new parameters $x$ and $y$ satisfying $z_1 = 1/ 4 x^2$ and $z_2 = y /4$, and then parameters $x_1$ and $x_2$ such that
$$
x_1 = (1-x),~~~~~x_2 = {\sqrt{y} \over 1-x }.
$$
The double scaling limit is given by letting $x_1 = \epsilon^2 u$ and $x_2 = 1/u$, and then sending $\epsilon \rightarrow 0$. Taking this limit in our $j$-function \jpone\ for the elliptic curve mirror to local $\IP^1 \times \IP^1$, we get
$$
j(\tau)= {64 (3+u^2)^3 \over (u^2-1)^2},
$$
which is indeed exactly the $j$-function of the $SU(2)$ Seiberg-Witten curve.

\newsec{Open Questions and Speculations}

In this paper we showed how to use symmetries to constrain the topological string amplitudes. As a result, we obtained nice expressions for the amplitudes in terms of (almost) holomorphic modular forms. However, various open questions remained, and new ideas for future research emerged.

\vskip 0.2cm
\noindent{{\it i. }{\it Compact case. }}Our formalism is completely
general, and applies to both compact and non-compact Calabi-Yau
threefolds. However, all the examples that we worked out explicitly
consisted in non-compact target spaces. As explained in section 4.1,
the reason is that in the compact case the period matrix $\tau_{IJ}$
does not have positive definite imaginary part. It would be
interesting to understand how to get modular expressions in this
case, perhaps using the closely related matrix ${\cal N}_{IJ}$, as
also explained in section 4.1. \vskip 0.2cm

\noindent{{\it ii. }{\it
Full group of symmetries. }}In this paper, we considered the group of
symmetries of the topological string generated by monodromies of the
periods. However, as explained in the introduction, this is just a
subgroup of the full group of symmetries, which consists in the group
of $\omega$-preserving diffeomorphisms. In the local case, the $\omega$ preserving diffeomorphisms were
used in \ih\ to solve completely the
topological string. It would be very interesting to see if this
generalizes to compact Calabi-Yau manifolds.
\vskip 0.2cm

\noindent{{\it iii. }{\it Away from the weak coupling. }}
In this work we obtained nice modular expressions for the topological
string amplitudes genus by genus. However, the main object of study
was the topological string wave function $Z(g_s, x)$,
which should make sense at any value of the string coupling.
It would be interesting to use the symmetries to constrain the
topological string amplitude for all values of the string coupling.
This would correspond to solving the equations \fund\ away from the weak coupling regime. However, this may be hard, as one has to pick the correct non-perturbative definition of \fund .

%%%%%%%%%%%%

\vskip 0.15 in

{\noindent \bf Acknowledgements}
\vskip 0.07 in

We would like to thank Jim Bryan, Tom Coates, Robbert
Dijkgraaf, Chuck Doran, Thomas Grimm, Minxin Huang, Marcos Mari\~no, Andrew Neitzke, Nikita Nekrasov, Yong-Bin
Ruan, Albert Schwarz, Jan Stienstra, Cumrun Vafa and Don Zagier for useful
discussions. We would also like to thank Ruza Markov for pointing out a few misprints in the first version of the paper. The research of M.A. is supported in part by a DOI OJI
Award, the Alfred P. Sloan Fellowship, and the NSF grant PHY-0457317. A.K. is supported in part by the DOE-FG02-95ER40896 grant.
V.B. is supported by an MSRI postdoctoral fellowship for the ``New
topological structures in physics" program, and by an NSERC
postdoctoral fellowship.

\appendix{A}{Modular Forms and Quasi-Modular Forms}

In this appendix we review essential facts in the theory of modular forms and quasi-modular forms, mainly in order to fix our conventions.

Denote by ${\cal H} = \{ \tau \in \IC | {\rm Im}(\tau) > 0\}$ the complex upper half-plane, and let $\Gamma \subset SL(2,\IZ)$ be a subgroup of finite index.

The action of the modular group $\Gamma$ on ${\cal H}$ is given by
$$
\tau \mapsto \frac{A \tau +B}{C \tau +D}, ~\rm{for}~ \gamma = \pmatrix{A&B\cr C&D} \in \Gamma.
$$

A {\it modular form} of weight $k$ on $\Gamma$ is a holomorphic function $f: {\cal H} \to \IC$ satisfying
$$
f( \gamma \tau) = (C \tau + D)^k f (\tau)~~~\rm{for~all}~\gamma = \pmatrix{A&B\cr C&D} \in \Gamma,
$$
and growing at most polynomially in $1/{\rm Im}(\tau)$ as ${\rm Im}(\tau) \to 0$.

We can also define an {\it almost holomorphic modular form} of weight $k$ on $\Gamma$ as a function ${\hat f}: {\cal H} \to \IC$ satisfying the same transformation property and growth condition as above, but with the form
$$
{\hat f}(\tau,{\bar \tau}) = \sum_{m=0}^{M} f_m (\tau) {\rm Im}(\tau)^{-m},
$$
for some integer $M \geq 0$, where the functions $f_m (\tau)$'s are holomorphic. The constant term in the series, $f_0 (\tau)$, is a {\it quasi-modular form} of weight $k$; it is holomorphic, but not quite modular. It has the form
$$
f_0(\tau) = \sum_{m=0}^{M} h_m (\tau) E_2(\tau)^{m},
$$
where the $h_m(\tau)$'s are holomorphic modular forms and we defined the second Eisenstein series
$$
E_2(\tau) =1 - 24 \sum_{n=1}^\infty {n q^n \over (1- q^n)},
$$
which is itself quasi-modular of weight $2$. Its almost holomorphic counterpart is defined as
$$
E_2^*(\tau, {\bar \tau}) = E_2 (\tau) - \frac{3}{\pi {\rm Im}(\tau)}.
$$
Note that there is an isomorphism between the ring of almost holomorphic modular forms and the ring of quasi-modular forms.

\subsec{Modular Forms of $\Gamma(2)$}

Our conventions for the theta functions with characteristics are as follows:
$$
\theta\left[{a\atop b}\right](z,\tau) =
\sum_n q^{{1\over 2}(n+a)^2} e^{2 \pi i (n+ a)(b+z)}.
$$
As usual, we denote the $\Gamma(2)$ theta constants by
$$\theta_2 = \theta\left[{1\over 2}\atop 0\right](0|\tau),\qquad
\theta_3 = \theta\left[0\atop 0\right](0|\tau),\qquad
\theta_4 = \theta\left[0\atop {1\over 2}\right](0|\tau)
$$
We also define the fourth powers
$$
b:=\theta_2^4(\tau),\qquad
c:=\theta_3^4(\tau),\qquad
d:=\theta_4^4(\tau),
$$
which satisfy the identity $c = b+d$. Also, $\eta^{12} = 2^{-4} b c d$, where $\eta$ is the Dedekind $\eta$-function.

Here are some useful formulae involving derivatives of modular forms:
$$24 q {d\over d q} \log(\eta) = E_2,$$
$$ 6 q{d  \over  d q} \log (d) = E_2 - b-c,$$
$$ 6 q{d  \over  d q} \log (c) = E_2 + b-d,$$
$$ 6 q{d  \over  d q} \log (b) = E_2 + c +d.$$

\subsec{Modular Forms of $\Gamma(3)$}

For the congruence subgroup $\Gamma(3)$, the relevant theta constants (taking their third powers) are\foot{We use the same variables to denote the fourth powers of the $\Gamma(2)$ theta constants and the third powers of the $\Gamma(3)$ theta constants, but it should always be clear from the context which subgroup we are considering.}
$$
a:=\theta^3\left[{1\over 6}\atop {1\over 6}\right](0,\t), \quad
b:=\theta^3\left[{1\over 6}\atop {1\over 2} \right](0,\t), \quad
c:=\theta^3\left[{1\over 6}\atop {5\over 6}\right](0,\t), \quad
d:=\theta^3\left[{1\over 2} \atop {1\over 6}\right](0,\t),
$$
satisfying the identities
$$
b=a+c, \qquad d= a+\alpha b,
$$
with $\alpha= e^{2 \pi i \over 3}$. Moreover, the Dedekind $\eta$-function is given by $\eta^{12} = {i \over 3^{3/2}} a b c d$.

We need derivative formulae for these theta constants as well. Let us first define the six following modular forms of weight $2$:
$$
\eqalign{
t_1 &= \frac{a c}{\eta^2},~~~~~t_2= \frac{a b}{\eta^2},~~~~~t_3 = \frac{b c}{\eta^2},\cr
~~~~~t_4&= \frac{b d}{\eta^2},~~~~~
t_5 = \frac{a d}{\eta^2},~~~~~t_6= \frac{c d}{\eta^2}.
}
$$
Then we found the relations:
$$
\eqalign{
8 q \frac{d}{dq} \log a &= {1\over 3} E_2 \left(\frac{\tau+1}{3} \right) = E_2(\tau) - \frac{2}{3} (t_4 + t_6 + \alpha t_3),\cr
8 q \frac{d}{dq} \log b &= {1 \over 3} E_2 \left(\frac{\tau}{3} \right) = E_2(\tau) + \frac{2}{3} (t_1 - t_5 + t_6),\cr
8 q \frac{d}{dq} \log c &= {1 \over 3} E_2 \left(\frac{\tau+2}{3} \right) = E_2(\tau) + \frac{2}{3} (t_4 + t_5 - \alpha^2 t_2),\cr
8 q \frac{d}{dq} \log d &= 3 E_2 (3 \tau ) = E_2(\tau) + \frac{2}{3} (-t_1 + \alpha^2 t_2 + \alpha t_3).}
$$
Note that the second equality in each line are `triple' analogs of the doubling identities for the Eisenstein series $E_2(\tau)$.

\appendix{B}{Siegel modular forms}

A good reference on Siegel modular forms is Ghitza's elementary introduction \Ghitza\ and the more complete textbook \Klingen.

Let $\Gamma$ be a subgroup of finite index of the symplectic group $Sp(2r,\IZ)$ defined by
$$
Sp(2r,\IZ) = \left \{ \pmatrix{ A&B\cr C&D}\in GL(2r,\IZ)| A^T C = C^T A, B^T D = D^T B, A^T D - C^T B = I \right \},
$$
where $I$ is the $r \times r$ identity matrix. Define the {\it Siegel upper half space}
$$
{\cal H}_r = \{\t \in \rm{Mat}_{r\times r} (\IC) | \t^T = \t, {\rm Im}(\t) > 0\};
$$
this is the space of $r \times r$ symmetric matrices with positive definite imaginary part. The action of $\Gamma$ on ${\cal H}_r$ is given by
$$
\t \mapsto (A \t+B)(C \t +D)^{-1}~~{\rm for}~~\gamma = \pmatrix{A&B\cr C&D} \in \Gamma.
$$

A weight $k$ {\it (scalar-valued) Siegel modular form} of $\Gamma$ is a holomorphic function $f: {\cal H}_r \to \IC$ satisfying
$$
f( \gamma \tau) = \det (C \tau + D)^k f (\tau)~~~\rm{for~all}~\gamma = \pmatrix{A&B\cr C&D} \in \Gamma.
$$
Note that for $r>1$ we do not need to impose the condition of
holomorphicity at infinity in the definition of a modular form, as was
the case for $r=1$.

Moreover, for $r>1$ one can define more general objects,
which transform under irreducible representations of $GL(r,\IC)$. Given such a representation $\rho:
GL(r,\IC)\rightarrow GL(V)$, where $V$ is a finite-dimensional vector space, we say that a function transforming under $\rho$ is a Siegel modular form of weight $\rho$ --- see for instance
\Ghitza.

We can also defined `generalized' theta functions as
$${\theta\left[a \atop b\right](z_i,\tau)=\sum_{n\in \IZ^r}\exp\left(\pi i
\sum_{ij} (n^i+a^i)\tau_{ij}(n^j+a^j)+ 2 \pi i \sum_i (z_i+b_i)n^i\right)\ , }
$$
where $a$, $b$ and $z$ are vectors of length $r$.

\listrefs

\end

%%%%%%%%%%%%%%%%%%%%%%%%%%%%%%%%%%%%%%%%%%%%%%

\newsec{A Quintic Example}

For definiteness, consider the quintic Calabi-Yau manifold
$X$, given as a hypersurface $f=0$ in ${\IP}^4,$ where
$$
f =\sum_{k=1}^5 x_k^5 - 5\psi \prod_{k=1}^5 x_k.
$$

This has a holomorphic $(3,0)$ form ${\Omega}$
$$
\Omega = 5 \psi {x_5 \;d x_1\wedge d x_2 \wedge d x_3 \over \del f/\del x_4}
$$
At least localy, ${\psi}$ parameterizes the moduli space of complex structures on $X$, which is one dimensional, so $b_3(X)=2$.

The symmetry group ${\Gamma}$ is generated by two elements, $M_{0}$ and $M_{1}$ (the notation will become clear momentarily).
The element $M_{0}$ satisfies $(M_0)^5=id$ and corresponds to
circling the ${\psi}=0$ point:
$$
M_{0}\;: \qquad {\psi} \rightarrow {\alpha}\, {\psi}
$$
where ${\alpha}^5= 1$.  These are discrete diffeomorphisms of the
theory: on the one hand, they preserve both ${\Omega}$ and $f$, and
can be undone by a coordinate transformation that takes $x_1
\rightarrow {\alpha}^{-1} x_1$, leaving the rest invariant.
Because of this, it is natural to use not $\psi$ but
$$
z = {1\over \psi^5}
$$
as a good coordinate on the moduli space.
While ${\Omega}$ is invariant under the $M_0$ action it's
periods are not. In terms of the natural basis of periods
at large radius
$$
{\Pi} =
\pmatrix{P_1 \cr P_0 \cr X^1 \cr X^0}
$$
circling ${\psi}=0$ is represented by an $Sp(4,\IZ)$ action:
$$
{\Pi}(\alpha \psi) = M_0 \, \Pi(\psi)
$$
where
$$
M_{0}=
\pmatrix{-19 &-3&5&-3\cr 31 &-4&-8&-5\cr -80&-11&21&-11\cr
0&1&0&1}.
$$
Above, $P_0$, $P_1$ correspond to $6$ and $4$ cycles of the mirror,
respectively, and $X^0$ and $X^1$ to $0$ and $2$-cycles\foot{In other words,
$(P_1 , P_0 ,X^1 , X^0) \sim X^0 (-{5\over 2}t^2,- {5\over 6} t^3, t,1),$ where $t=X^1/X^0.$},

The second generator $M_1$ corresponds to monodromy around $\psi =1$,
the conifold point, where $B_0$ is a vanishing cycle.
Correspondingly, any cycle which intersects it is ill defined -- this
is just the Picard-Lefshetz monodromy. This corresponds to the
monodromy matrix
$$
M_1= \pmatrix{1 &0&0&0\cr 0 &1&0&0\cr 0&0&1&0\cr
0&-1&0&1}.
$$
In the $\psi$ plane, there are four more copies of this, corresponding
at $\psi = \alpha^{k},\; k=1,\ldots 4,$ which all correpsond to the same point in the $z$ plane. We'll mainly use the $z-$plane perspective.

By holomorphy, there must be a third singular point in the $z$-plane.
This corresponds to the large radius point, $\psi=\infty$, with monodromy
$$
M_{\infty} = (M_0 M_1)^{-1}.
$$

To summarize: we can view the moduli space as $\rm{\IP}^1$, with three points $z=0,
{1},\infty$ deleted,
$$
{\cal M}/\Gamma = {\IP}^1/\{0,{1},\infty\}.
$$
Given any symplectic basis of $H_3(X)$, the
corresponding periods of $\Omega$ become sections of a flat
$Sp(4,\IZ)$ bundle this, with monodromies $M_0$,$M_1$
and $M_\infty,$ up to conjugation.